\documentclass[journal]{IEEEtran}

\usepackage{cite}
\usepackage{url}

\usepackage[pdftex]{graphicx}
\usepackage{xcolor}
\usepackage{cuted}

\usepackage{amsmath}
\usepackage{amssymb}
\usepackage{booktabs}
\usepackage{nicematrix }

\usepackage{array}
\usepackage{tabularx}

\ifCLASSOPTIONcompsoc
  \usepackage[caption=false,font=normalsize,labelfont=sf,textfont=sf]{subfig}
\else
  \usepackage[caption=false,font=footnotesize]{subfig}
\fi

\usepackage{mathtools}

\def\tensor#1{\mathord{\buildrel{\lower3pt\hbox{$\scriptscriptstyle\leftrightarrow$}} 
\over{#1}}} 

\mathchardef\lm="2D
\newcommand{\lp}{\scalebox{0.9}{+}}

\usepackage[normalem]{ulem} 
\usepackage{color}
\def\edit#1{{\color{red} #1}} 

\def\edit#1{#1} 
\def\sout#1{} 
\begin{document}
\title{The Swiss Roll Metamaterial Revisited}

\author{Zachary~Fritts$^1$
        and~Anthony~Grbic$^2$
\thanks{Z.F. gratefully acknowledges the support of the Rackham Predoctoral Fellowship from the University of Michigan.

$^1$ Department of Mechanical and Aerospace Engineering, University of California, Irvine, CA, 92697 USA; e-mail: zfritts@uci.edu.

$^2$Department of Electrical and Computer Engineering, University of Michigan, Ann Arbor, MI, 48109 USA; e-mail: agrbic@umich.edu. }
}

\maketitle

\begin{abstract}
In this work, we analytically model the Swiss roll metamaterial medium, showing that each unit cell can be thought of as a flux-coupled waveguide terminated in coupled inductors. The equivalent circuit model of the metamaterial accurately predicts its response, including its higher order resonances, even when made of a lossy, finite-thickness conductor. A closed-form solution for the response of this circuit model is then derived by solving a matrix equation of arbitrarily large dimensions. Along the way, we provide a general method for modeling the effective permeability of uniaxial, anisotropic metamaterials formed from axially invariant conducting inclusions.
\end{abstract}

\IEEEpeerreviewmaketitle

\section{Introduction}
\label{sec:Introduction}
\IEEEPARstart{J}{ust} over 25 years have passed since Pendry and his coauthors showed that the Swiss roll metamaterial enables the realization of negative effective permeabilities at RF frequencies \cite{1999pendry_SwissRoll}. A few years prior, the possibility of negative permittivity from a wire medium -- a metamaterial plasma -- had also been theorized and experimentally demonstrated \edit{\cite{1953brown_IndexLessThan1,model1955propagation,1955brown_ArtificialDielectricsCM,1962rotman_ArtificialPlasma,1996pendry_NegEpsWireMedium,1998pendry_WirePlasmaExpt}}. This set the stage for the first experimental demonstration of a negative-index metamaterial \cite{2000smith_DoubleNegative}, which had been theoretically predicted by Viktor Veselago \cite{1968veselago_DoubleNegative}. Since Pendry had argued that a negative-index electromagnetic medium would enable perfect lensing \cite{2000pendry_NegIndPerfectLens}, a flurry of research ensued, resulting in the first experimental demonstration of imaging beyond the diffraction limit in 2004 \cite{2004grbic_DiffractionLimit}. 

Major themes of later research in electromagnetic metamaterials were anticipated in the 1999 publication of negative permeability metamaterials \cite{1999pendry_SwissRoll}. A key component of the first negative-index metamaterial was a two-dimensional cousin of the Swiss roll metamaterial: the split-ring resonator (SRR). SRRs were described in 1952 by Schelkunoff and Friis \cite{1952schelkunoff_AntennasTheoryPractice}, were fabricated at least by 1981 \cite{1981hardy_SRRs}, and were treated as a homogeneous metamaterial in \cite{1999pendry_SwissRoll}. Additionally, the authors mention the potential inclusion of nonlinear materials within resonant metamaterials, since the field-enhancement due to resonance would significantly increase nonlinear effects. A modified version of the Swiss roll structure was an early example of a chiral metamaterial \cite{2009demetriadou_ChiralSwissRoll}. Nonlinear and chiral metamaterial media are presently active areas of research in metamaterials.

The Swiss roll structure itself had a number of experimental incarnations, first as a resonator for enhancing the RF magnetic field that causes coherent precession of the magnetic moments of hydrogen nuclei in magnetic resonance imaging (MRI) \cite{2001wiltshire_SwissRollMRI}, then as negative permeability metamaterials \cite{2003wiltshire_SwissRollSeminar,2004wiltshire_1DArraySwissRolls}, and even as a metamaterial absorber \cite{2015chen_SwissRollAbsorber}. 

\begin{figure}[!t]
\centering
    \centering
    \subfloat[ ]{\includegraphics[width=0.5\columnwidth]{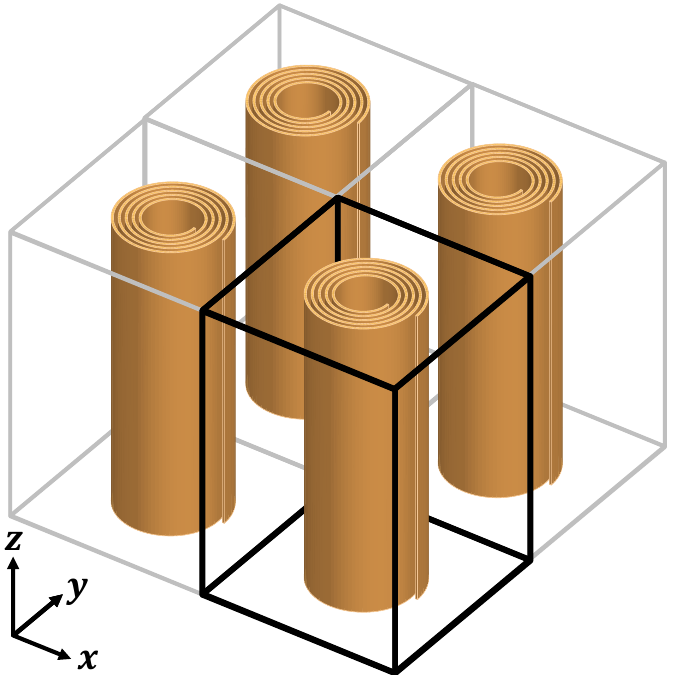}}
    \hfill
    \subfloat[ ]{\includegraphics[width=0.45\columnwidth]{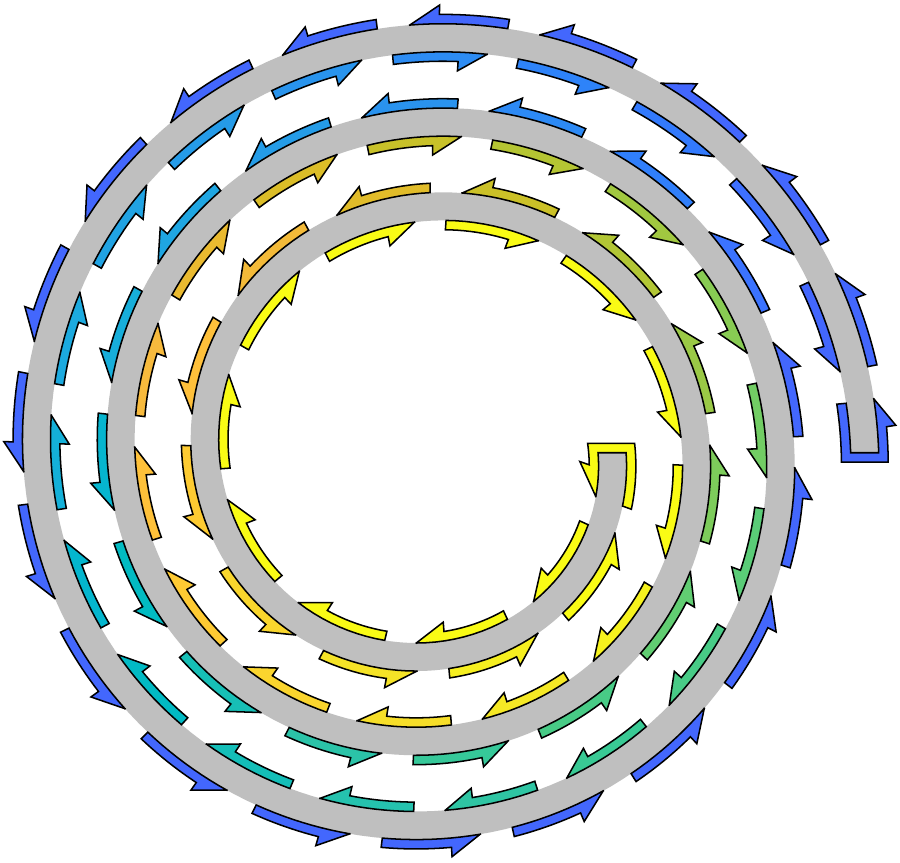}}
    \\
    \subfloat[ ]{\includegraphics[width=0.48\columnwidth]{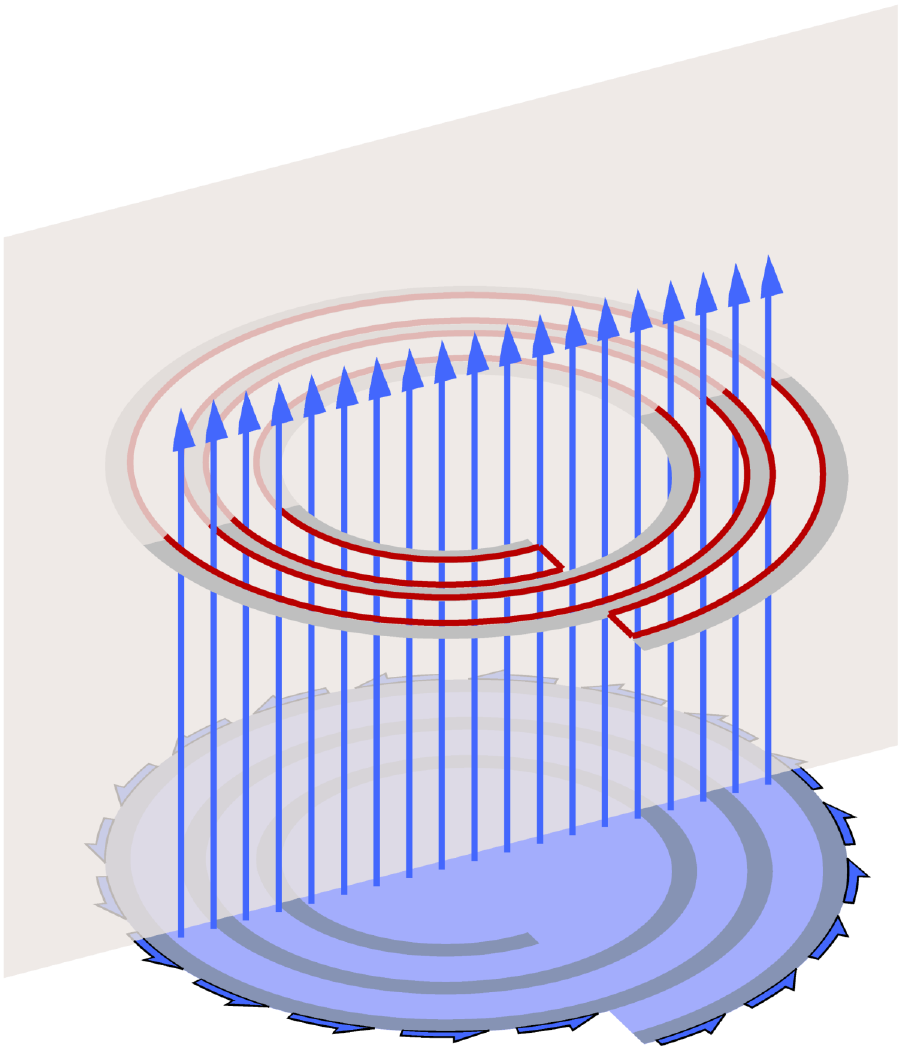}}
    \hfill
    \subfloat[ ]{\includegraphics[width=0.48\columnwidth]{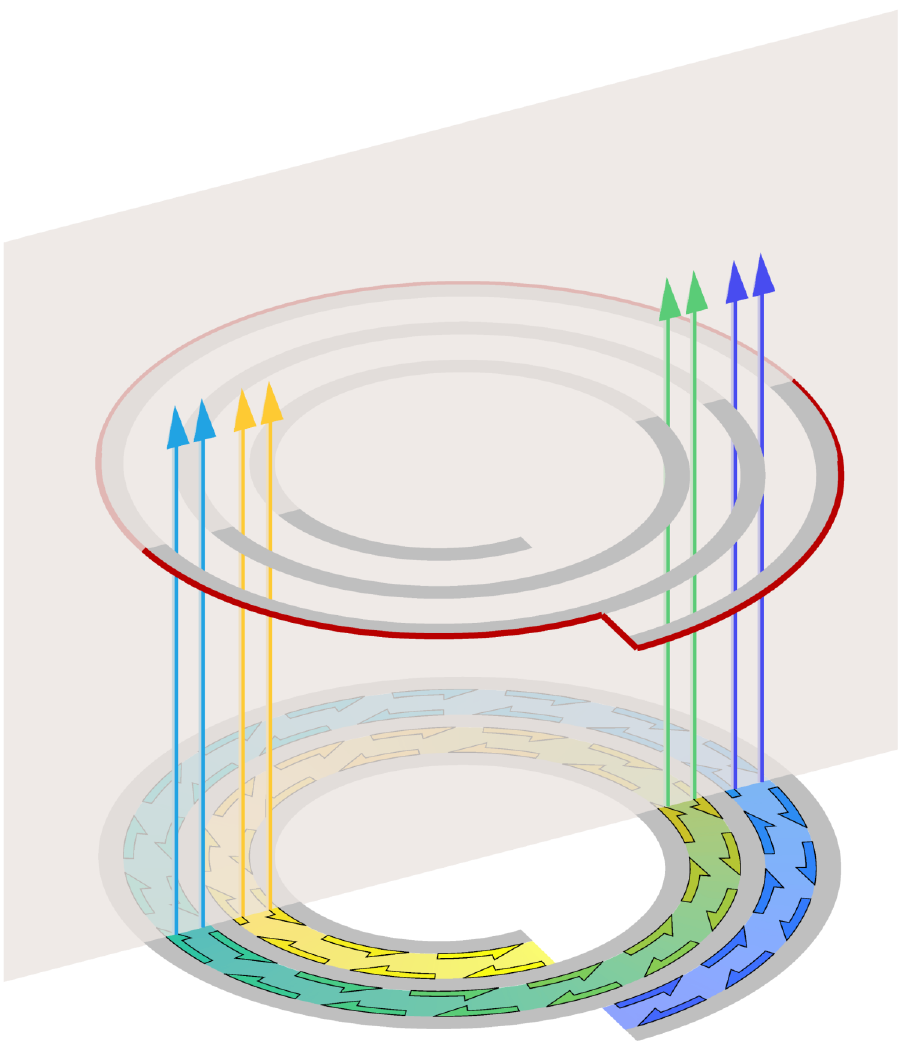}}
\caption{The Swiss roll metamaterial\edit{\sout {medium}}. (a) Four unit cells of a Swiss roll metamaterial. (b) An example current distribution on a cross section of the Swiss roll with finite-thickness conductor. \edit{\sout{showing how the} Note how} current wraps around the ends of the conductor \edit{\sout{to continue its flow}}. The innermost \edit{\sout{circular region} arc of the spiral} (yellow arrows) \edit{\sout{can be thought of as} forms} an inductor \edit{\sout{, as can the region bounded by the current flowing on the exterior of the Swiss roll}}. The gap between adjacent \edit{\sout{layers} turns} of the conductor forms a waveguide. (c) Current flowing on the exterior of the Swiss roll produces an axial magnetic field (\edit{blue} shading and \edit{vertical} arrows) that pierces \edit{\sout{the closed loop (outlined in red) that bounds}} the waveguide \edit{(outlined in red).} \edit{\sout{showing} This shows} that flux can couple into the waveguide. (d) Similarly, current in the waveguide produces a magnetic field (shaded) whose flux lines pierce the closed loop (outlined in red) that bounds the exterior of the Swiss roll.}
\label{fig:SR_4Cells}
\end{figure}

Despite the apparent simplicity of the Swiss roll structure (Fig. \ref{fig:SR_4Cells}a), the electromagnetic models of its behavior have been essentially heuristic. The original model was derived by analogy with a circuit formed of a multi-turn inductor and series capacitors, assuming zero-thickness conductors \cite{1999pendry_SwissRoll}. Simulations of the Swiss roll metamaterial revealed that it supports a number of higher-order modes, \textit{even in the metamaterial limit}, when the size of the unit cells is a small fraction of a wavelength \cite{2007zolla_SwissRollSims,2009demetriadou_SwissRollSims}. These higher order modes were found to closely align with the resonant frequencies of an open-ended section of parallel-plate waveguide \cite{2007zolla_SwissRollSims}, but no model of the Swiss roll metamaterial was found that predicted these resonances.

As a tribute to the pioneering work of Pendry and his predecessors, we revisit the Swiss roll metamaterial and derive a circuit model for its effective permeability from first principles. Our circuit model neatly predicts the higher order modes of the Swiss roll metamaterial and is valid even for lossy, finite-thickness conductors. Interestingly, our work clarifies that the Swiss roll should not be thought of as a multi-turn inductor. Rather, it can be thought of as coupled inductors which terminate the ends of a flux-coupled transmission line. This flux-coupled transmission line is so called because magnetic flux within the transmission line mutually couples to currents flowing around the Swiss roll. To illustrate this, Fig. \ref{fig:SR_4Cells}c shows how currents on the exterior of the Swiss roll produce a magnetic field that penetrates the spiral of the Swiss roll. This spiral acts as a waveguide, and the current distribution in the guide conversely produces a magnetic field with mutual flux through the exterior perimeter of the Swiss roll (Fig. \ref{fig:SR_4Cells}d).

Finding the electromagnetic behavior of this system involves solving a high-dimensional system of equations, which (perhaps surprisingly) admits an analytic solution.  

\section{Homogenizing the Permeability: A Circuit Perspective}
\label{sec:HomogenizingByCirc}
\subsection{Effective Permeability from the Impedance of a Unit Cell}
\label{subsec:RelatingMuAndZ}
The Swiss roll metamaterial is formed of a regular two-dimensional lattice of unit cells containing a conductive sheet that has been wound into a spiral cylinder around the $z$-axis (Fig. \ref{fig:SR_4Cells}). It is assumed that the unit cells are electrically small within the lattice plane, i.e., $w_{c}\ll \lambda$, and that each unit cell is infinitely long along the axis of the cylinder. (This assumption is not necessary, but it will simplify the mathematics appreciably.) \edit{\sout{In fact,}} The Swiss roll metamaterial is part of a larger class of metamaterials formed of two-dimensional lattices of axially-invariant unit cells, including wire media where the conductors are arranged parallel to a common axis. Like all such metamaterials, the effective material properties are anisotropic and naturally separate into axial and transverse components. \edit{Most\sout{The majority}} of this article is aimed at providing an accurate model of the axial permeability, $\mu_{zz}$, of the Swiss roll metamaterial, since this is the component of the permeability tensor where all the interesting behavior occurs. 

Given any 2D metamaterial lattice of conductive inclusions, finding the effective axial permeability $\mu_{zz}$ of this metamaterial can be cast as finding the effective inductance of a unit cell. If we could associate an ``input impedance" $Z_{in}$ with the response of the unit cell, then the effective inductance of this unit cell would simply be: 
\begin{equation}
    \label{eqn:Leff}
    L_{eff}=\frac{Z_{in}}{j\omega}.
\end{equation}
How should the impedance $Z_{in}$ be defined such that the associated effective inductance $L_{eff}$ yields the axial permeability of the metamaterial? Begin by selecting a single unit cell (such as the unit cell highlighted in black in Fig. 1), and breaking up the axial magnetic field in the unit cell $H_{z}$ into local and scattered components,
\begin{equation}
    \label{eqn:H_Loc_Scat}
    H_{z}=H_{z}^{loc}+H_{z}^{scat},
\end{equation}
where the local field consists of the incident field and the scattered field from all other unit cells in the metamaterial. Since the unit cell is electrically small, the incident field in the cell is approximately uniform. If the scattered magnetic field from adjacent unit cells is approximately uniform, it follows that the local magnetic field inside of the unit cell is approximately uniform also. 

\begin{figure}[!t]
    \centering
    \subfloat[  ]{\includegraphics[width=0.54\columnwidth]{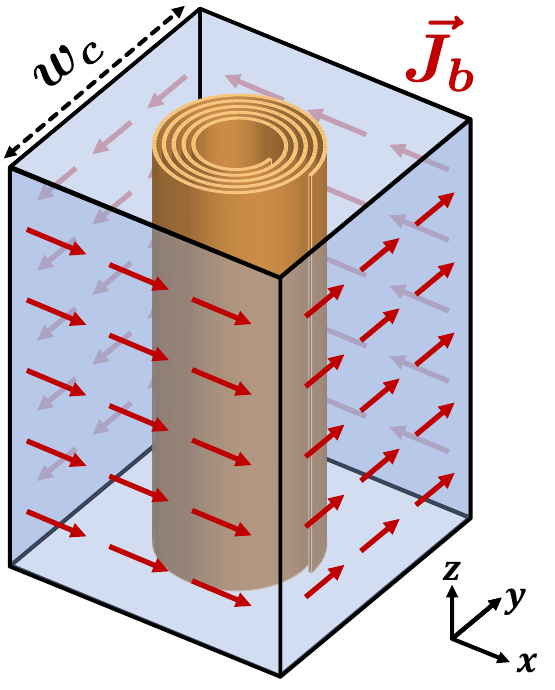}}
    \centering
    \subfloat[  ]{\includegraphics[width=0.46\columnwidth]{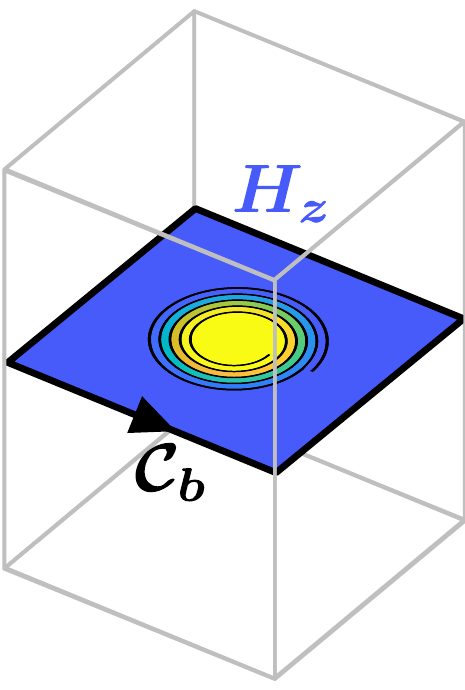}}
\caption{Isolating a single unit cell with the equivalence principle. (a) The local magnetic field in the unit cell is generated by equivalent electric current $\vec{J}_b$ on the boundary of the unit cell. (b) Conduction currents in the unit cell cause the total field to be inhomogeneous, altering the total magnetic flux through the surface bounded by closed loop $\mathcal{C}_b$.}
\label{fig:EquivalencePrinciple}
\end{figure}

Now, according to Love's equivalence theorem \cite{2001harrington_EquivalencePrinciple}, we can replace all sources and scatterers external to the chosen unit cell by empty space if we place electric and magnetic currents on the boundary of the unit cell that regenerate the local field in the unit cell. The requisite electric current is given by: 
\begin{equation}
    \label{eqn:J_EquivalencePrinc}
    \vec{J}_{b}\coloneqq J_b(\hat{z}\times \hat{n})=H_{z}^{loc}(\hat{z}\times \hat{n}),
\end{equation}
where $\hat{n}$ is the outward-facing normal vector to a point on the boundary of the unit cell, so that $\vec{J}_{b}$ is directed along the boundary of the cell in the transverse plane. On the hypothesis that the local magnetic field is uniform throughout the unit cell, it is easy to see from (\ref{eqn:J_EquivalencePrinc}) that the source current is uniform and transverse to the axis of the cylinder, as illustrated in Fig. \ref{fig:EquivalencePrinciple}. Since the unit cell is electrically small, this uniform shell of electric current around the boundary of the unit cell is alone responsible for generating the local magnetic field in the unit cell; the magnetic current required by the equivalence principle does not contribute in the quasistatic limit.

The local magnetic field induces electric currents on the conducting inclusion within the unit cell, and these currents generate the scattered field $H_{z}^{scat}$. This scattered field is, in general, inhomogeneous, as illustrated in Fig. \ref{fig:EquivalencePrinciple}\edit{\sout{, altering the total axial magnetic field -- and hence the total axial magnetic flux}}. By Faraday's law, the electromotive force (EMF) around the boundary of the cell is related to the \edit{total} axial magnetic flux by: 
\begin{equation}
    \label{eqn:Faraday_Fig2}
    \mathcal{E}_{b}=-j\omega \Phi_{z},
\end{equation}
where the EMF and the flux are given by the integrals: 
\begin{align}
    \label{eqn:Fig2_EMF_Flux_Def}
    \mathcal{E}_{b} & \coloneqq \oint_{\mathcal{C}_b}\vec{E}\cdot \mathrm{d}\vec{\ell} \\
    \Phi_{z} & \coloneqq \iint_{\mathcal{S}_{b}}B_{z}\mathrm{d}s.
\end{align}
We can now define the ``input impedance" of the unit cell as the ratio of the induced EMF and the current around the boundary of the unit cell: 
\begin{equation}
    \label{eqn:Zin_Definition}
    Z_{in}\coloneqq -\frac{\mathcal{E}_{b}}{J_{b}}.
\end{equation}
Since $J_{b}$ is uniform around the boundary of the cell, there is no ambiguity in this definition of the impedance $Z_{in}$. 

To relate the impedance $Z_{in}$ to the effective permeability, note that in a unit cell filled with a homogeneous medium of permeability $\mu_{zz}$, there is no scattered field. Consequently, the EMF around the unit cell is given by: 
\begin{equation}
    \label{eqn:EMF_EmptyCell}
    \mathcal{E}_{b}=-j\omega\iint_{\mathcal{S}_{b}}B_{z}\mathrm{d}s=-j\omega (\mu_{zz}\mu_{0}w_{c}^{2})H_{z}^{loc},
\end{equation}
where $w_{c}^{2}$ is the cross-sectional area of the unit cell. Since $J_{b}=H_{z}^{loc}$, the impedance of the homogeneous unit cell is simply: 
\begin{equation}
    Z_{in}=j\omega(\mu_{zz}\mu_{0}w_{c}^{2}).
\end{equation}
In light of (\ref{eqn:Leff}), we can therefore write the relative permeability in terms of the effective inductance as: 
\begin{equation}
    \label{eqn:MuEff_Definition}
    \boxed{ \mu_{zz}=\frac{L_{eff}}{\mu_{0}w_{c}^{2}}=\frac{Z_{in}}{j\omega(\mu_{0}w_{c}^{2})}. } 
\end{equation}
Applying this equation to any unit cell, not just one filled with a homogeneous medium, results in an effective permeability for that unit cell. Note that $\mu_{0}w_{c}^{2}$ represents the effective inductance of a unit cell containing only free space, so that the relative permeability can be viewed as a ratio of inductances. The integration required to find the total flux through the unit cell represents the a kind of field averaging, satisfying the intuition that effective material parameters arise from an averaging process \cite{2006smith_FieldAveraging}.

\subsection{A General Method for Finding $Z_{in}$}
\label{subsec:GeneralMethod}
Since we can write the effective permeability in terms of the impedance $Z_{in}$, finding this impedance is the key to homogenizing the metamaterial. The general procedure for finding this impedance is as follows:

\begin{enumerate}
    \item assume a reasonable form for the currents and magnetic fields inside of the unit cell, where the boundary current $J_{b}$ is the only ``known" current
    \item form $N$ independent closed loops in the unit cell, where $N$ is the number of independent currents
    \item apply Faraday's law in integral form $N$ times around these independent closed loops, hence forming $N$ equations for the unknown currents in terms of the EMFs around the closed loops
    \item enforce boundary conditions by determining the EMFs
    \item (optional) interpret the resulting equations using an equivalent circuit
    \item solve the linear system for the input impedance
    \item relate the input impedance to the effective permeability using (\ref{eqn:MuEff_Definition}).
\end{enumerate}

In order to illustrate the full process, we apply it in the next section to a simple example.

As a matter of terminology, we use ``electromotive force" or ``EMF" to denote the line integral of the electric field along a given curve, open or closed. The symbol $\mathcal{E}$ is reserved for such electric line integrals, with subscripts indicating the appropriate curve. Closed curves are all denoted by the symbol $\mathcal{C}$, while open curves or line segments are all denoted by $s$ and distinguished by the superscript $s$. Examples of this notation include: 
\begin{equation}
\label{eqn:EMF_Notation}
    \mathcal{E}_o=\oint_{\mathcal{C}_o}\vec{E}\cdot\mathrm{d}\vec{\ell}, \quad \mathcal{E}_1^s=\int_{s_1}\vec{E}\cdot\mathrm{d}\vec{\ell}.
\end{equation}

\section{Metamaterial Media of Conducting Circular Cylinders}
\label{sec:Circular_Cylinders}
\begin{figure}[!t]
    \centering
    \subfloat[  ]{\includegraphics[width=0.45\columnwidth]{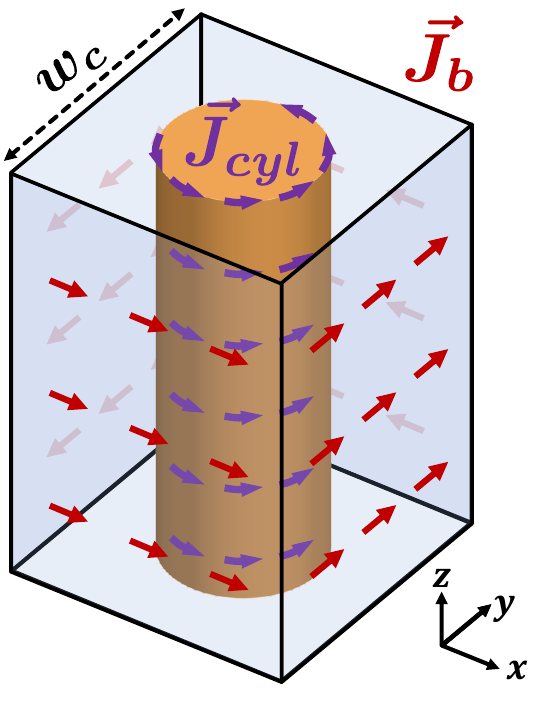}}
    \centering
    \subfloat[  ]{\includegraphics[width=0.55\columnwidth]{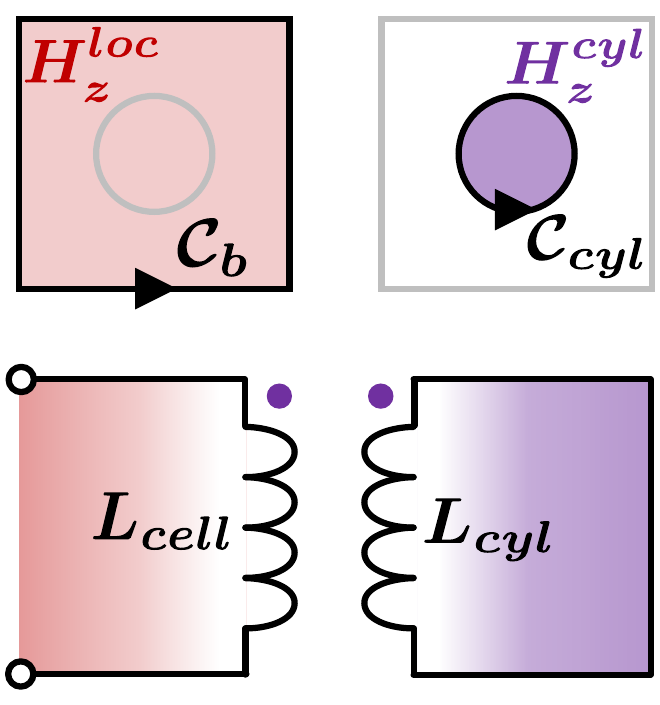}}
\caption{Currents and \edit{magnetic} fields for a unit cell containing a PEC cylinder. (a) Currents $\vec{J}_b$ (red arrows) on the boundary of the unit cell and $\vec{J}_{cyl}$ (purple arrows) on the surface of the cylinder are uniformly distributed if the cell width $w_c$ is electrically small. (b) The current $\vec{J}_b$ and $\vec{J}_{cyl}$ generate uniform axial magnetic fields throughout the unit cell and the cross-section of the cylinder, respectively. Applying Faraday's law around the regions $\mathcal{C}_b$ and $\mathcal{C}_{cyl}$ yields an equivalent circuit model consisting of coupled inductors.  }
\label{fig:PEC_Cylinder}
\end{figure}

Among the simplest examples of conducting scatterers that a unit cell could contain are conducting circular cylinders (Fig. \ref{fig:PEC_Cylinder}). By treating this unit cell using the method outlined \edit{\sout{above}} in Sec. \ref{subsec:GeneralMethod}, we hope to illustrate each of the steps involved. Since \cite{1999pendry_SwissRoll} also analyzes this simple unit cell as a first example, the methods can be compared. To demonstrate how to apply \edit{the} proposed method to lossy, finite thickness conductors, a unit cell containing an annular conducting cylinder is also analyzed in this section.

\subsection{Permeability of a Unit Cell Containing a Solid PEC Cylinder}
\label{subsec:PEC_Cylinder}
Consider a unit cell containing a circular PEC cylinder of radius $a$ (Fig. \ref{fig:PEC_Cylinder}). The the unit cell is of width $w_{c}$ and \edit{\sout{it}} is assumed to be invariant along the axial direction. There are two electric currents to account for in the unit cell. The first of these is the current $\vec{J}_{b}$ on the boundary of the unit cell which acts as the equivalent source of the axial component of the local magnetic field, $H_{z}^{loc}$, as related by (\ref{eqn:J_EquivalencePrinc}). The second electric current, $\vec{J}_{cyl}$ is the current induced on the surface of the PEC cylinder. Given that the radius of the cylinder is much smaller than a wavelength, this current will be azimuthally invariant to a very good approximation, so that we can write: 
\begin{equation}
    \label{eqn:Jcyl}
    \vec{J}_{cyl}(\rho,\phi)=J_{cyl}\delta(\rho-a)\hat{\phi},
\end{equation}
where $\hat{\phi}$ is the unit vector in the azimuthal direction. In the quasistatic limit, the magnetic field associated with the current $\vec{J}_{cyl}$ is entirely interior to the cylinder, and is uniformly distributed across this region. Therefore, we can write: 
\begin{equation}
    \label{eqn:Hcyl}
    \vec{H}_{cyl}=
\begin{cases}
J_{cyl}\hat{z}, & \rho< a \\
0,  & \rho>a. 
\end{cases}
\end{equation}
Equations (\ref{eqn:J_EquivalencePrinc}) and (\ref{eqn:Jcyl})-(\ref{eqn:Hcyl}) constitute step 1 of the method detailed in Section (\ref{subsec:GeneralMethod}). 

Given the assumed forms of the magnetic fields in the cell, the only contribution to the magnetic field outside of the PEC cylinder is due to the local field. This is different than the analysis in \cite{1999pendry_SwissRoll}, where the unit cell is assumed to be long, but of finite length. If the cylinder is assumed to be of finite length, a depolarization field must be included to account for the return of the magnetic flux passing through the interior of the cylinder. Despite the fact that our analysis does not include this field, it will be seen shortly that assuming infinite-length unit cells yields the same effective permeability as assuming a uniformly distributed depolarization field. 

Determining the behavior of the metamaterial involves applying Faraday's law in integral form, so closed loops must be chosen over which the integration will be applied. Since there are two independent electric currents, two independent closed loops will be chosen. The natural choices for these closed loops are the loop traversing the boundary of the cell, $\mathcal{C}_{b}$, and that around the circumference of the PEC cylinder, $\mathcal{C}_{cyl}$, as shown in Fig. \ref{fig:PEC_Cylinder}. Both closed loops are oriented following the right-hand rule around the $z$-axis.

Assuming that each of the currents exist within free space, the integral form of Faraday's law is simply
\begin{equation}
    \label{eqn:FaradayGeneral}
    \oint_{\mathcal{C}} \vec{E}\cdot \mathrm{d}\vec{\ell}=-j\omega \mu_{0}\iint_{\mathcal{S}} H_{z}\mathrm{d}s,
\end{equation}
where $\mathcal{C}$ is a closed loop in the transverse plane and $\mathcal{S}$ is the planar surface bounded by said curve. Applied to the curve $\mathcal{C}_{b}$ around the boundary of the unit cell, this yields: 
\begin{equation}
    \label{eqn:PEC_Cyl_EMF_b}
    \mathcal{E}_{b}=-j\omega(\mu_{0}w_{c}^{2})J_{b}-j\omega(\mu_{0}\pi a^{2})J_{cyl},
\end{equation}
where $\mathcal{E}_b$ is the EMF around the boundary $\mathcal{C}_b$ of the unit cell. Similarly, applying (\ref{eqn:FaradayGeneral}) around the circumference of the PEC cylinder gives:
\begin{equation}
    \label{eqn:PEC_Cyl_EMF_cyl}
    \mathcal{E}_{cyl}=-j\omega(\mu_{0}\pi a^{2})J_{b}-j\omega(\mu_{0}\pi a^{2})J_{cyl},
\end{equation}
where $\mathcal{E}_{cyl}$ is the EMF around the cylinder,
\begin{equation}
    \label{eqn:PEC_Cyl_EMF_LoopInt}
    \mathcal{E}_{cyl} \coloneqq \oint_{\mathcal{C}_{cyl}}\vec{E}\cdot \mathrm{d}\vec{\ell}.
\end{equation}
This completes steps 2-3 of the procedure outlined in Section \ref{subsec:GeneralMethod}. Up until this point, we have not used the fact that the cylinder is perfectly conducting. Enforcing this boundary condition simply involves noting that $\mathcal{E}_{cyl}=0$, as the tangential electric field along the boundary of the cylinder must vanish. Consequently, (\ref{eqn:PEC_Cyl_EMF_b}) and (\ref{eqn:PEC_Cyl_EMF_cyl}) yield the following coupled circuit equations: 
\begin{equation}
    \label{eqn:PEC_Cyl_Zmatrix}
    \begin{bmatrix}
    -\mathcal{E}_{b} \\
    0
    \end{bmatrix}
    =
    \begin{bmatrix}
    j\omega L_{cell}  & j\omega L_{cyl} \\
    j\omega L_{cyl} & j\omega L_{cyl}
    \end{bmatrix}
    \begin{bmatrix}
    J_{b} \\
    J_{cyl}
    \end{bmatrix},
\end{equation}
where the self and mutual inductances are simply proportional to the relevant cross-sectional areas through which magnetic flux passes,
\begin{equation}
    \label{eqn:PEC_Cyl_CircParams}
    L_{cell}\coloneqq \mu_{0}w_{c}^{2}, \quad L_{cyl}\coloneqq \mu_{0}\pi a^{2}.
\end{equation}
Note that (\ref{eqn:PEC_Cyl_Zmatrix}) is the set of circuit equations for an inductor $L_{cell}$ that is magnetically coupled to a short-circuited inductor $L_{cyl}$, as shown in Fig. \ref{fig:PEC_Cylinder}. Since the inductor $L_{cyl}$ is short-circuited, no flux can pass through it, which is consistent with the fact that perfect conductors exclude magnetic fields from their interior. \edit{\sout{Incidentally, the circuit model for this unit cell also represents a strategy for protecting sensitive electronics in wireless power transfer}} 

Solving the system (\ref{eqn:PEC_Cyl_Zmatrix}) for the input impedance gives,
\begin{equation}
    \label{eqn:PEC_Cyl_Zin}
    Z_{in}=-\frac{\mathcal{E}_{b}}{J_{b}}=j\omega(L_{cell}-L_{cyl}),
\end{equation}
so that the effect of coupling the cell inductance to the short-circuited inductance of the cylinder is to reduce the total inductance. This makes sense, as the overall effect of the PEC cylinder \edit{\sout{here}} is to reduce the total magnetic flux flowing through the unit cell. This is reflected in the effective permeability of the homogenized metamaterial as found from (\ref{eqn:MuEff_Definition}), 
\begin{equation}
    \label{eqn:PEC_Cyl_muEff}
    \mu_{zz}=\frac{Z_{in}}{j\omega L_{cell}}=1-\frac{L_{cyl}}{L_{cell}}=1-\frac{\pi a^{2}}{w_{c}^{2}},
\end{equation}
showing that the relative permeability is reduced from unity by the fill fraction $F\coloneqq \frac{A_{cyl}}{A_{cell}}$. The same conclusion is reached if one includes a uniform depolarization current, as in \cite{1999pendry_SwissRoll}.

\subsection{Permeability of a Unit Cell Containing a Lossy, Annular Conductor}
\label{subsec:AnnularCyl}
To demonstrate how to apply the proposed method to finite thickness, lossy conductors, we next analyze a unit cell containing an annular cylinder. The outer radius of the annulus is $a_{o}$, and its inner radius is $a_{i}$, while its conductivity $\sigma$ is modeled by the tangential boundary condition: 
\begin{equation}
    \label{eqn:Lossy_BC}
    \vec{J}_{\parallel}=\eta_{s}\vec{E}_{\parallel},
\end{equation}
where the sheet impedance $\eta_{s}$ is given in terms of the conductivity $\sigma$ or the skin depth $\delta$ by: 
\begin{equation}
    \label{eqn:Sheet_EtaS_Def}
    \eta_{s}=\sqrt{ \frac{\omega \mu}{2\sigma} }\left( 1+j \right)=\frac{\omega \mu \delta}{2}(1+j).
\end{equation}
If the conductor is much thicker than the skin depth, the inductive portion of the sheet impedance is vanishingly small compared to other inductances that appear in the model, so that the sheet impedance can be regarded as purely resistive. 

The assumed forms of the electric currents and their associated magnetic fields are largely the same as in Section \ref{subsec:PEC_Cylinder}, except for the addition of a current $\vec{J}_{i}$ that can flow on the inner surface of the annulus. The magnetic field distributions associated with the azimuthally-invariant electric currents flowing on the inner and outer surfaces of the conductor take the form:
\begin{align}
    \label{eqn:Ann_Cyl_Hfields}
    \vec{H}_{i} & = 
    \begin{cases}
    J_{i}\hat{z} & \rho<a_{i} \\
    0 & \rho>a_{i}
    \end{cases} \\
    \vec{H}_{o} & = \begin{cases}
    J_{o}\hat{z} & \rho<a_{o} \\
    0 & \rho>a_{o}.
    \end{cases}
\end{align}

\begin{figure}[!t]
    \centering
    \subfloat[  ]{\includegraphics[width=0.33\columnwidth]{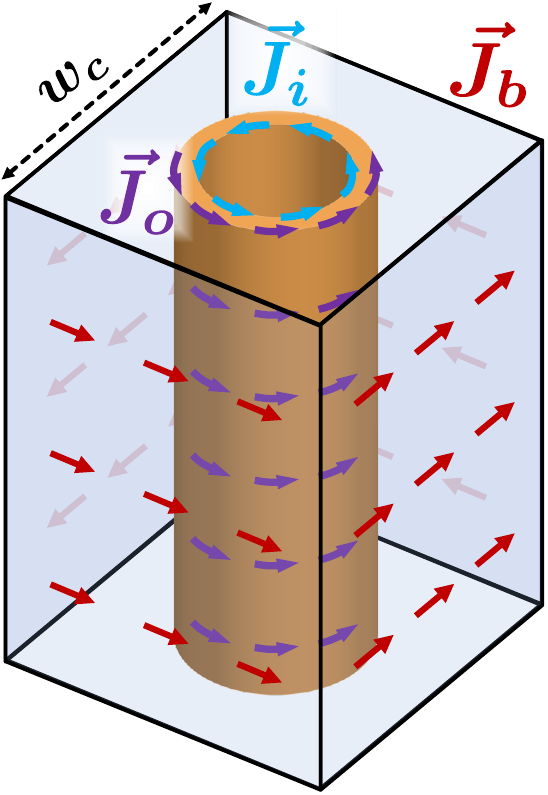}}
    \centering
    \subfloat[  ]{\includegraphics[width=0.67\columnwidth]{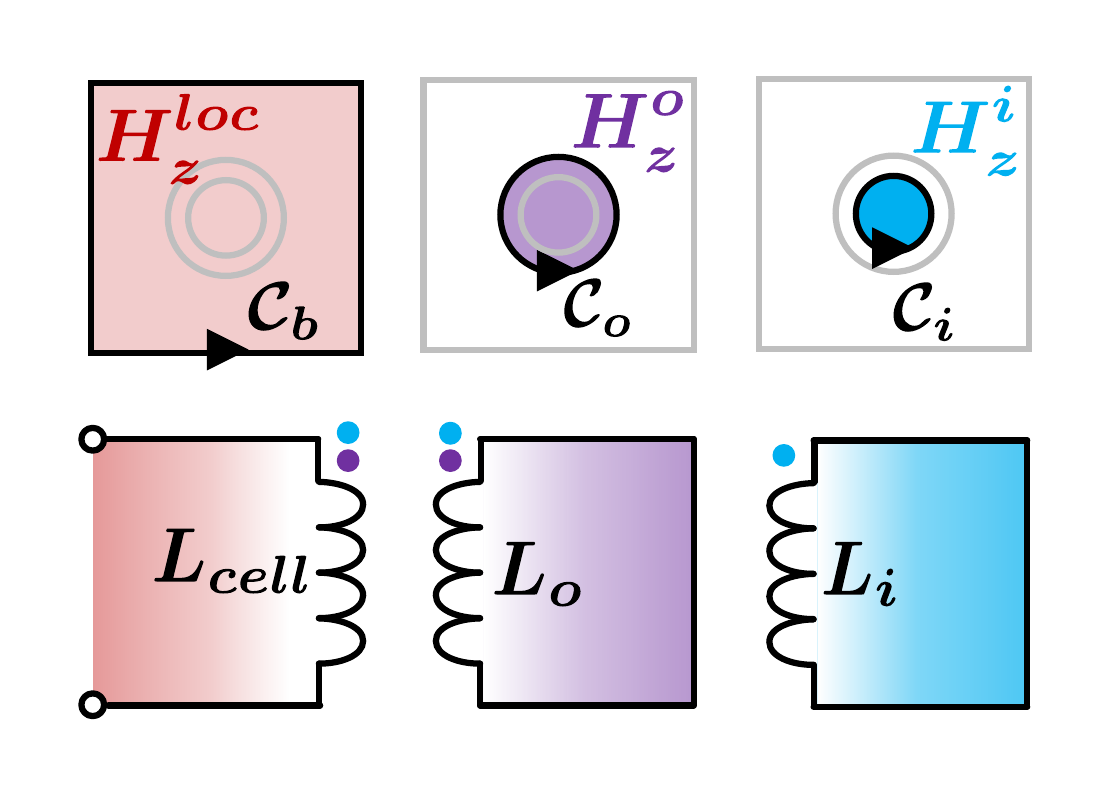}}
\caption{Currents and \edit{magnetic} fields for a unit cell containing an annular conducting cylinder. (a) Currents $\vec{J}_b$ (red arrows) on the boundary of the unit cell, $\vec{J}_{o}$ (purple arrows) on the outer surface of the annulus, and $\vec{J}_{i}$ (purple arrows) on the inner surface of the annulus are uniformly distributed if the cell width $w_c$ is electrically small. (b) The currents $\vec{J}_b$, $\vec{J}_{o}$, and $\vec{J}_{i}$ generate uniform axial magnetic fields throughout the unit cell and the cross-section of the cylinder, respectively. Applying Faraday's law around the regions $\mathcal{C}_b$, $\mathcal{C}_{o}$, and $\mathcal{C}_{i}$ yields an equivalent circuit model consisting of three coupled inductors. \edit{Multiple dots over an inductor signify multiple independent mutual inductances.} Dots of the same color indicate inductors coupled by the same mutual inductance.}
\label{fig:Annular_Cyl}
\end{figure}

In addition to applying Faraday's law around the perimeter of the unit cell, $\mathcal{C}_{b}$, it is natural to apply it around the outer circumference $\mathcal{C}_{o}$ and inner circumference $\mathcal{C}_{i}$ of the annular conductor. These closed loops, the assumed electric current densities, and their associated magnetic field distributions are illustrated in Fig. \ref{fig:Annular_Cyl}.

Application of Faraday's law in integral form along the indicated closed loops yields the equations: 
\begin{align}
\label{eqn:Ann_Cyl_EMF}
    -\mathcal{E}_{b} & =j\omega(\mu_{0}w_{c}^{2})J_{b}+j\omega(\mu_{0}\pi a_{o}^{2})J_{o}+j\omega(\mu_{0}\pi a_{i}^{2})J_{i} \nonumber \\
    -\mathcal{E}_{o} & =j\omega(\mu_{0}a_{o}^{2})J_{b}+j\omega(\mu_{0}\pi a_{o}^{2})J_{o}+j\omega(\mu_{0}\pi a_{i}^{2})J_{i} \\
    -\mathcal{E}_{i} & =j\omega(\mu_{0}a_{i}^{2})J_{b}+j\omega(\mu_{0}\pi a_{i}^{2})J_{o}+j\omega(\mu_{0}\pi a_{i}^{2})J_{i}, \nonumber
\end{align}
\edit{\sout{where the EMFs have the form
\begin{equation}
    \label{eqn:Ann_Cyl_EMF_integral}
    \mathcal{E}_{\alpha} \coloneqq \oint_{\mathcal{C_\alpha}}\vec{E}\cdot \mathrm{d}\vec{\ell}.
\end{equation}}}
Substituting the tangential boundary condition (\ref{eqn:Lossy_BC}) into the above equation yields expressions for the EMFs associated with the inner and outer boundary of the annular conductor in terms of the associated electric currents: 
\begin{equation}
    \label{eqn:Ann_Cyl_EMF_BC}
    \mathcal{E}_{o,i}=\eta_{s}(2\pi a_{o,i})J_{o,i}.
\end{equation}
Consequently, the full system of circuit equations for the unit cell can be written in matrix form as: 
\begin{equation}
    \label{eqn:Ann_Cyl_Zmatrix}
    \begin{bmatrix}
    -\mathcal{E}_{b} \\
    0 \\
    0
    \end{bmatrix}
    =
    \left[
    \begin{NiceArray}{c|cc}
    j\omega L_{c} & j\omega L_{o} & j\omega L_{i} \\
    \hline \\\\[-3.5\medskipamount]
    j\omega L_{o} & j\omega L_{o}+R_{o} & j\omega L_{i} \\
    j\omega L_{i} & j\omega L_{i} & j\omega L_{i}+R_{i}
    \end{NiceArray}
    \right]
    \begin{bmatrix}
    J_{b} \\
    J_{o} \\
    J_{i}
    \end{bmatrix},
\end{equation}
where the associated circuit parameters are given by: 
\begin{align}
    \label{eqn:Ann_Cyl_CircParams}
    L_{c} & =\mu_{0}w_{c}^{2} \nonumber \\
    L_{o,i} & = \mu_{0}\pi a_{o,i}^{2} \\
    R_{o,i} & = (2\pi a_{o,i}) \eta_{s}. \nonumber 
\end{align}
These circuit equations are those of three (lossy) coupled inductors, as shown in Fig. \ref{fig:Annular_Cyl}. This system of equations has the following block matrix form:
\begin{equation}
    \label{eqn:Block_Zmatrix_3x3}
    \begin{bmatrix}
    V_{b} \\
    0
    \end{bmatrix}
    =
    \begin{bmatrix}
    Z_{11} & \mathbf{Z}_{12} \\
    \mathbf{Z}_{21} & \mathbf{Z}_{22}
    \end{bmatrix}
    \begin{bmatrix}
    I_{1} \\
    \mathbf{I}_{2}
\end{bmatrix},
\end{equation}
whose formal solution is given by the Schur complement as
\begin{equation}
    \label{eqn:Schur_Soln_3x3_General}
    Z_{in} = \frac{V_b}{I_1} =  Z_{11} - \mathbf{Z}_{12}\mathbf{Z}_{22}^{-1}\mathbf{Z}_{21},
\end{equation}
which will appear repeatedly in the following sections.

The solution to the input impedance associated with the circuit equations (\ref{eqn:Ann_Cyl_Zmatrix}) as found using (\ref{eqn:Schur_Soln_3x3_General}) is given by: 
\begin{equation*}
    \label{eqn:Ann_Cyl_Zin}
    Z_{in}=j\omega L_{c}-\frac{j\omega L_{o}(j\omega L_{i}\lp R_{i})\lp j\omega L_{i}R_{i}\lp  \frac{(j\omega L_{i})^{2}(R_{o}\lp R_{i})}{j\omega(L_{o}\lm L_{i})}}{(j\omega L_{i}\lp R_{i}) \lp  \frac{L_{i}(R_{o}\lp R_{i})}{L_{o}\lm L_{i}}\lp \frac{R_{i}R_{o}}{j\omega (L_{o}\lm L_{i})}}.
\end{equation*}
Under the assumption that the conductor is much thicker than the skin depth, the equation for the input impedance simplifies significantly. Under that assumption, the quantity $\frac{R_{o}+R_{i}}{j\omega(L_{o}-L_{i})}$ is vanishingly small, so applying a small argument approximation of the input impedance over this value yields: 
\begin{equation}
    \label{eqn:Ann_Cyl_Zin_Approx}
    Z_{in}\approx j\omega(L_{c}-L_{o})+R_{o}.
\end{equation}
Note that assuming that the conductor was much thicker than the skin depth resulted in an impedance that depends only on the size of the cell and on the outer diameter of the annular conductor -- the inner conductor does not matter. This makes sense, as the outer surface of the conductor shields the inner surface from the incident wave, just as in coaxial cables or in shielded resonant loops \cite{2014grbic_ShieldedLoopRes}. The corresponding effective permeability is simply: 
\begin{equation}
    \label{eqn:Ann_Cyl_muEff}
    \mu_{zz} \approx \left( 1-\frac{L_{i}}{L_{c}} \right)-\frac{R_{o}}{j\omega L_{c}} = \left( 1-\frac{\pi a_{o}^{2}}{w_{c}^{2}} \right)+j\pi\frac{\delta a_{o}}{w_{c}^{2}},
\end{equation}
where $\delta$ is the skin depth, and the small inductance associated with the skin depth has been neglected. Due to the effects of shielding, this is the same relative permeability as obtained for the circular PEC cylinder in the prior section, but with an added loss term due to the finite conductivity of the conductor.

\section{Swiss Roll Metamaterial with an Electrically Short Spiral}
\label{sec:ShortSR}
\subsection{Deriving the Cell Impedance when $d\ll \lambda$}
\label{subsec:Zin_Short_SR}

\renewcommand\tabularxcolumn[1]{m{#1}}
\begin{table}[b]
\renewcommand{\arraystretch}{2.0}
\caption{Subregion Properties for the Swiss Roll}
\label{table:SubregionProperties}
\centering
\begin{tabularx}{1.00\columnwidth}
{|>{\centering\arraybackslash\hsize=0.9\hsize}X
|>{\centering\arraybackslash\hsize=0.9\hsize}X
|>{\centering\arraybackslash\hsize=0.9\hsize}X
|>{\centering\arraybackslash\hsize=0.9\hsize}X
|>{\centering\arraybackslash\hsize=1.4\hsize}X|}
\hline
\textbf{Boundary Contour} & \textbf{Current} & \textbf{Axial Magnetic Field} & \textbf{Cross Sectional Area} & \textbf{Inductance} \\
\hline
$\mathcal{C}_b$  & $J_b$  & $H_z^{loc}$  & $w_c^2$  & $L_b\coloneqq\mu_0 w_c^2$  \\
\hline
$\mathcal{C}_o$  & $J_o$  & $H_z^{o}$  & $A_o$  & $L_o\coloneqq\mu_0A_o$  \\
\hline
$\mathcal{C}_i$  & $J_i$  & $H_z^{i}$  & $A_i$  & $L_i\coloneqq\mu_0A_i$  \\
\hline
$\mathcal{C}_1$  & $J_1$  & $H_z^{1}$  & $\frac{gd}{2}$  & $\frac{L'd}{2}=\frac{\mu_0gd}{2}$  \\
\hline
$\mathcal{C}_2$  & $J_2$  & $H_z^{2}$  & $\frac{gd}{2}$  & $\frac{L'd}{2}=\frac{\mu_0gd}{2}$  \\
\hline
\end{tabularx}
\end{table}

Having illustrated our general approach on simple examples, we can consider the Swiss roll unit cell. The conductor of the Swiss roll \edit{(thickness $t$)} is wrapped $\mathcal{N}$ times around the axis of the unit cell. \edit{The inner and outer radii of the Swiss roll are $a_i$ and $a_o$, respectively, and there is} a uniform gap of width $g$ between adjacent turns. There must be at least one full turn for any gap to exist, so we require $\mathcal{N}>1$. However, the number of turns need not be an integer. One way of viewing the Swiss roll is as a parallel plate waveguide (PPW) wrapped around itself multiple times. In this section, we consider the case where the length $d$ of that PPW is electrically short, i.e., $d\ll \lambda$.

The \edit{\sout{PPW}}\edit{spiral waveguide} is characterized by per-unit-length inductance, capacitance, and (series) resistance: 
\begin{equation}
    \label{eqn:PPW_RLC_Params}
    \begin{split}
        L' & \coloneqq \mu_{0}g \\
        C' & \coloneqq \frac{\epsilon}{g} \\
        R' & \coloneqq 2\eta_{s},
    \end{split}
\end{equation}
where $\eta_{s}$ is defined as in (\ref{eqn:Sheet_EtaS_Def}) and the factor of two in the per-unit-length resistance accounts for loss from both conductors of the waveguide. Note that the curvature of the guide is \edit{\sout{here}} neglected for simplicity. 

Circuit models of transmission lines \edit{are} frequently \edit{\sout{are}} written as cascaded circuits containing both series and parallel impedances. The per-unit-length parameters (\ref{eqn:PPW_RLC_Params}) are associated with the following series and parallel impedances, normalized by unit length: 
\begin{equation}
    \label{eqn:PPW_Zs_Zp_Defs}
    Z_{s}\coloneqq j\omega L'+R', \quad Z_{p}\coloneqq \frac{1}{j\omega C'}.
\end{equation}
As will be seen shortly, the waveguide is flux-coupled to currents outside of the waveguide. The per-unit-length mutual impedance that describes this coupling is:
\begin{equation}
    \label{eqn:PPW_ZL_def}
    Z_{L}\coloneqq j\omega L'.
\end{equation}

\begin{figure}[t]
    \centering
    \subfloat[ ]{\includegraphics[width=0.8\columnwidth]{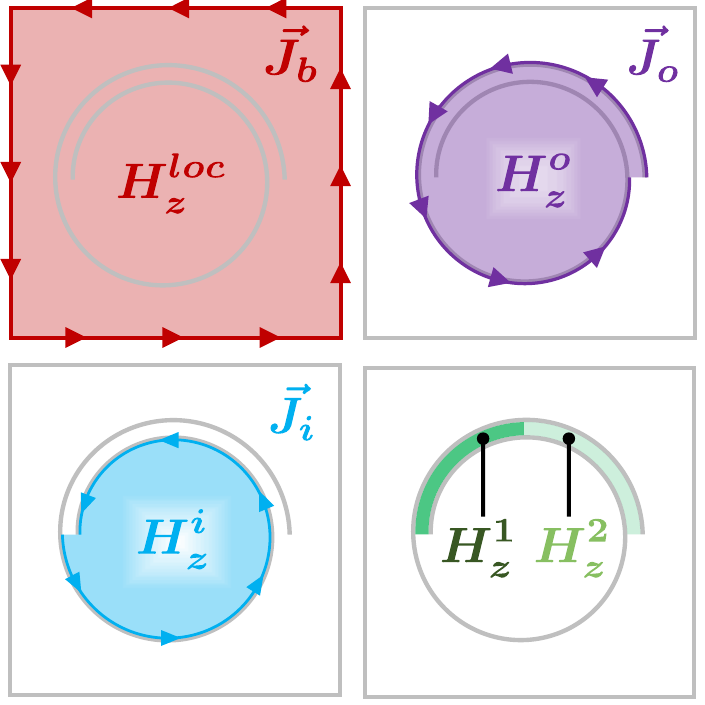}}
    \label{fig:_5a}
    \subfloat[ ]{\includegraphics[width=0.82\columnwidth]{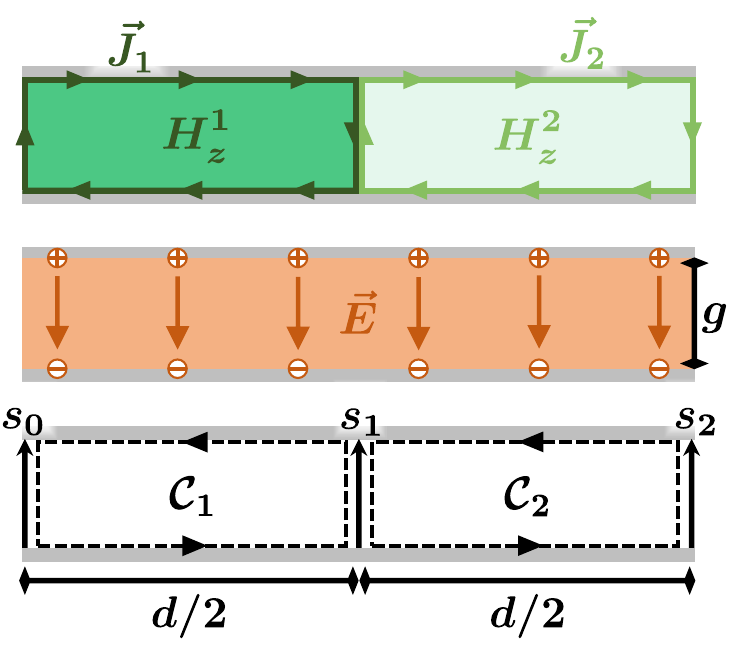}}
    \label{fig:_5b}
    \caption{Assumed field and current distributions for the Swiss roll unit cell with an \edit{electrically short spiral waveguide}. (a) The unit cell is broken up into five subregions (shaded), with a uniform current distribution conformed to the boundary of each subregion. Each current distribution produces a uniform axial magnetic field within its subregion. (b) Straightened waveguide subregions of the Swiss roll form sections of parallel-plate waveguide (PPW) of total length $d$. If the currents in each half of the PPW are not equal, then charge accumulates on the top and bottom conductors, producing an electric field inside the guide, which is assumed to be uniform.}
    \label{fig:Short_SwissRoll}
\end{figure}

To begin, let us break up the cross-section of the unit cell into five regions, as indicated in Fig. \ref{fig:Short_SwissRoll}. \textbf{Each region is associated with a uniformly distributed electric current on its boundary that produces a uniform, axial magnetic field within that region.} The largest of these regions is that covering the entire unit cell, which is bounded by current ${J}_{b}$. As in the case of the annular cylinder, there are also regions fitted to the outer and inner boundaries of the Swiss roll, carrying currents ${J}_{o}$ and ${J}_{i}$, respectively. Additionally, there are closed loops of current ${J}_{1}$ and ${J}_{2}$ that each circulate around half of the PPW. Note that the current distributions ${J}_{i}$ and ${J}_{1}$ each jump across the gap at the interior end of the PPW, while ${J}_{o}$ and ${J}_{2}$ each jump across the gap at the exterior end of the PPW. Physically, of course, no electric current exists in the gaps. Rather, by enforcing continuity of current at the ends of the PPW by the equations
\begin{equation}
    \label{eqn:KCL_Short_SR}
    J_{1}=-J_{i}, \quad J_{2}=J_{o},
\end{equation}
we ensure that the total \edit{conduction} current \edit{\sout{distribution}} on the Swiss roll is non-zero only on the surface of the conductor. \edit{Table \ref{table:SubregionProperties} summarizes the geometric and electromagnetic parameters associated with each of the five regions in Fig. \ref{fig:Short_SwissRoll}.} \edit{\sout{The correspondence between the electric currents, the closed loops whose paths they follow, the axial magnetic fields that they generate interior to said loops, and the cross-sectional area bounded by the loops is summarized in Table \ref{table:SubregionProperties}.}} All lengths and areas appearing in circuit parameters are written in Appendix \ref{Appendix_Geometry} in terms of the geometric parameters describing the Swiss roll unit cell.

Next, we apply Faraday's law in integral form (\ref{eqn:FaradayGeneral}) to each of the five regions. \edit{\sout{The main difference between our prior examples of Faraday's law and the present case is in the calculation of the EMF by the electric line integral, since there are now sections of these line integrals that are not adjacent to a conductor.}} Consider the EMF around $\mathcal{C}_{1}$, the first half of the PPW, as shown in Fig. \ref{fig:Short_SwissRoll}a. The electric line integral can be broken into contributions along the conductors and along the line segments $s_{0}$ and $s_{1}$: 
\begin{equation}
    \label{eqn:ShortSR_EMF_1_oint}
    \mathcal{E}_1=-\frac{R'd}{2}  J_{1}-\mathcal{E}_{0}^{s}+\mathcal{E}_{1}^{s}.
\end{equation}
where $\mathcal{E}_{k}^{s}$ is the EMF along line segment $s_{k}$, following the notation explained in (\ref{eqn:EMF_Notation}). Similarly, the EMF around the second half of the PPW can be written as
\begin{equation}
    \label{eqn:ShortSR_EMF_2_oint}
    \mathcal{E}_2=-\frac{R'd}{2}J_{2}-\mathcal{E}_{1}^{s}+\mathcal{E}_{2}^{s}.
\end{equation}
The line integral of the electric field across the half-way point of the waveguide, $\mathcal{E}_{1}^{s}$, appears in both (\ref{eqn:ShortSR_EMF_1_oint}) and (\ref{eqn:ShortSR_EMF_2_oint}) above. It is possible to express this EMF in terms of the currents $J_{1,2}$ by application of Gauss's law. To see this, note that if these two current\edit{s} are not equal, a charge imbalance $Q$ will appear between the conductors of the waveguide, where 
\begin{equation}
    \label{eqn:ShortSR_Q}
    Q=\frac{1}{j\omega}\left( J_{1}-J_{2} \right)
\end{equation}
by conservation of charge. If we assume that this charge is uniformly distributed across the inner and outer conductors of the waveguide (a very good approximation if the guide is electrically short), then a uniform electric field $\vec{E}$ appears between these conductors, as shown in Fig. \ref{fig:Short_SwissRoll}. The electric field can be related to the charge by application of Gauss's law in integral form. Since the electric field is assumed to be uniform, Gauss's law can be written by inspection, and the electric field is given by: 
\begin{equation}
    \label{eqn:ShortSR_Evec}
    \vec{E}=\frac{J_{1}-J_{2}}{j\omega \epsilon d} \hat{e}=\frac{Z_{p}}{gd}(J_{1}-J_{2})\hat{e},
\end{equation}
where $\epsilon$ is the permittivity of the dielectric within the Swiss roll, and $\hat{e}$ is a unit vector directed from the outer conductor of the guide to the inner conductor.

Now that the electric field $\vec{E}$ in the interior of the waveguide is known, the EMF between the conductors of the waveguide can be computed, yielding
\begin{equation}
    \label{eqn:ShortSR_EMF1}
    \mathcal{E}_{1}^{s}=-\frac{Z_{p}}{d}(J_{1}-J_{2}).
\end{equation}

Up to this point, we have only dealt with the electric \edit{field} line integrals \edit{of} Faraday's law. The magnetic flux \edit{surface} integrals \edit{of Faraday's law} are straightforward, since the magnetic field distributions are all uniform: 
\begin{equation}
    \label{eqn:ShortSR_FluxInt}
    \Phi_{1,2}=\mu_{0}\iint_{\mathcal{S}_{1,2}}H_{z}\mathrm{d}s=\frac{\mu_{0}gd}{2}\left( H_{z}^{loc}+H_{z}^{o}-H_{z}^{1,2} \right),
\end{equation}
where $\mathcal{S}_{1,2}$ are the surfaces bounded by the closed curves $\mathcal{C}_{1,2}$. Note that the magnetic field in the interior region of the Swiss roll $H_{z}^{i}$ does not contribute to the magnetic flux inside of the guide. Writing the axial magnetic fields in terms of their associated current distributions, Faraday's law requires:
\begin{equation}
    \label{eqn:ShortSR_Faraday_C12}
    -\mathcal{E}_{1,2}=\frac{Z_{L}d}{2}\left( J_{b}+J_{o}-J_{1,2} \right).
\end{equation}
Substituting (\ref{eqn:ShortSR_EMF_1_oint}), (\ref{eqn:ShortSR_EMF_2_oint}), and (\ref{eqn:ShortSR_EMF1}) into the above yields the circuit equations for the loops $\mathcal{C}_{1}$ and $\mathcal{C}_{2}$: 
\begin{align}
    \mathcal{E}_{0}^{s} & =\frac{Z_{L}d}{2}J_{b} + \left(\frac{Z_{L}d}{2} + \frac{Z_{p}}{d} \right)J_{o} + \left( \frac{Z_{s}d}{2} + \frac{Z_{p}}{d} \right)J_{i} \label{eqn:ShortSR_EMF0} \\ 
    \mathcal{E}_{2}^{s} & =-\frac{Z_{L}d}{2}J_{b} + \left( \eta_{s}d + \frac{Z_{p}}{d} \right)J_{o} + \frac{Z_{p}}{d}J_{i}, \label{eqn:ShortSR_EMF2} 
\end{align}
where we have applied the current continuity equations (\ref{eqn:KCL_Short_SR}) to reduce the number of independent currents. 

Applying Faraday's law to the loops around the unit cell boundary $(\mathcal{C}_{b})$, outer perimeter of the Swiss roll $(\mathcal{C}_{o})$, and inner perimeter of the Swiss roll $(\mathcal{C}_{i})$ yields the respective circuit equations: 
\begin{align}
    -\mathcal{E}_{c} & = j\omega L_{c}J_{b}+j\omega L_{o}'J_{o} + j\omega L_{i}'J_{i} \nonumber \\ 
    0  & = j\omega L_{o}J_{b}+ \left( j\omega L_{o}'+R_{o} \right)J_{o} + j\omega L_{i}'J_{i}+\mathcal{E}_{2}^{s} \label{eqn:ShortSR_EMF_System}\\
    0  & = j\omega L_{i}J_{b}+j\omega L_{i}J_{o}+(j\omega L_{i}+R_{i})J_{i}+\mathcal{E}_{0}^{s}, \nonumber
\end{align}
where the inductances are proportional to the cross-sectional areas of the associated regions, and the resistances are proportional to the length of the conductor adjacent to said regions, as summarized in Table \ref{table:SubregionProperties}. The primed inductances are similar to the unprimed inductances, but increased or decreased by half of the inductance of the waveguide:
\begin{equation}
    \label{eqn:ShortSR_Lprime_Def}
    L_{i}'=L_{i}+\frac{gd}{2}, \quad L_{o}'=L_{o}-\frac{gd}{2}.
\end{equation}
The primed inductances arise because enforcing the current continuity equations (\ref{eqn:KCL_Short_SR}) causes the current $J_{i}$ to enclose a total area that is the sum of $A_{i}$ and half the cross-sectional area of the guide. Similarly, the current $J_{o}$ encloses a total area that is $A_{o}$ minus half the area enclosed by the guide.

\begin{figure*}
    \centering
    \subfloat[ ]{\includegraphics[width=2.04\columnwidth]{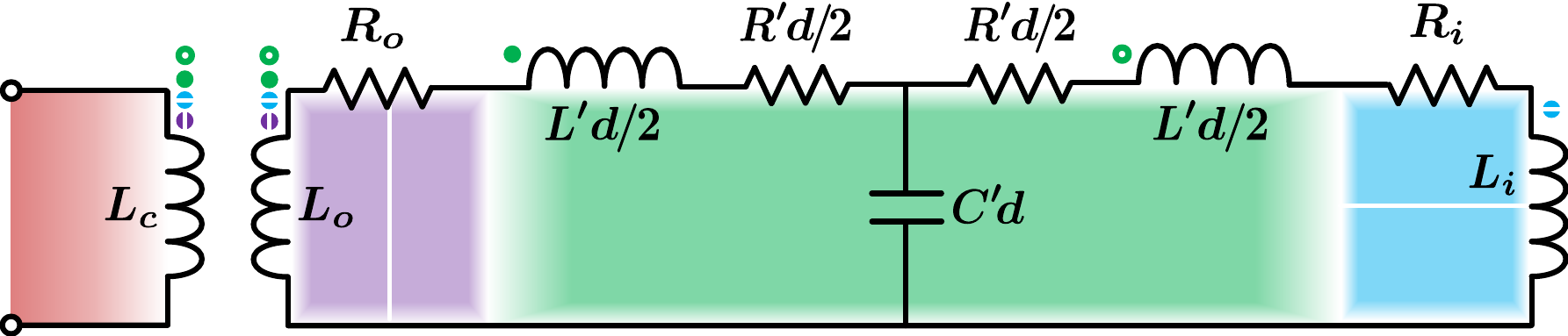}%
    \label{fig:Schematic_SR_Short_A}}

    \vspace{-1.0em} 

    \subfloat[]{\includegraphics[width=1.20\columnwidth]{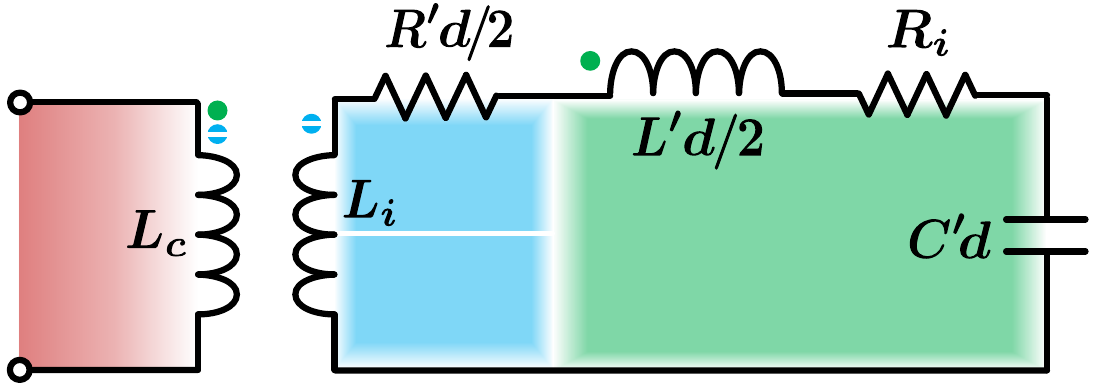}%
    \label{fig:Schematic_SR_Short_B}}
    \hfill
    \subfloat[]{\includegraphics[width=0.65\columnwidth]{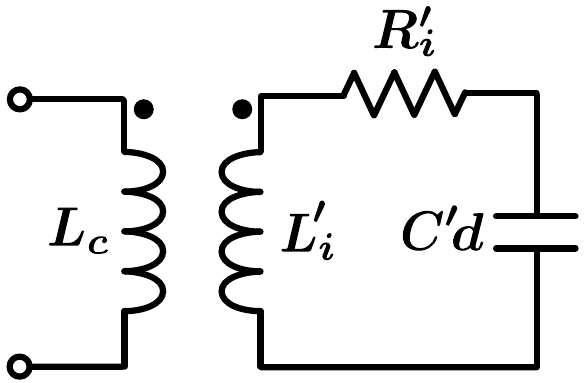}%
    \label{fig:Schematic_SR_Short_C}}
    \caption{Circuit models for the Swiss roll unit cell \edit{when the length $d$ of the spiral waveguide formed by the gap between the conductors is electrically short}. (a) The circuit model corresponding to the circuit equations (\ref{eqn:ShortSR_Zmatrix}) constructed given the current and field distributions shown in Fig. \ref{fig:Short_SwissRoll}, with shaded colors in the schematic corresponding to different subregions of the Swiss roll in that figure. (b) \edit{\sout{Except at frequencies far from resonance, the majority of the current flows through inductor $L_i$.}} Neglecting the current flowing on the outside of the Swiss roll by replacing $L_o$ with an open circuit yields \edit{this} simplified circuit model. \edit{The inductance $L_c$ of the empty unit cell couples inductively to the interior of the Swiss roll ($L_i$) and to the spiral waveguide. Since $d\ll \lambda$, the spiral waveguide has a lumped inductance and capacitance, $L'd/2$ and $C'd$. The effective permeability of the Swiss roll is negative for frequencies just above the (series) RLC resonance of the circuit.} \edit{This simplified circuit model is a good approximation to (a) except far from resonance.} (c) \edit{The final simplified circuit model of the Swiss roll when $d\ll \lambda$. The input impedance of this circuit corresponds to (\ref{eqn:ShortSR_Zin_Approx}).} \edit{\sout{Combining the series elements in (b) yields this schematic, whose input impedance corresponds to (\ref{eqn:ShortSR_Zin_Approx}).}}}
    \label{fig:SR_Short_Circuit}
\end{figure*}

Combining (\ref{eqn:ShortSR_EMF_System}) with (\ref{eqn:ShortSR_EMF0}) and (\ref{eqn:ShortSR_EMF2}), the full system of equations for the Swiss roll unit cell can be written in matrix form as: 
\begin{equation}
\label{eqn:ShortSR_Zmatrix}
    \begin{bmatrix}
    \lm \mathcal{E}_{c} \\
    0 \\
    0
    \end{bmatrix}
    =
    \left[
    \begin{NiceArray}{c|cc}
        j\omega L_{c} & j\omega L_{o}' & j\omega L_{i}' \\
        \hline \\\\[-3.5\medskipamount]
        j\omega L_{o}' & j\omega L_{o}' \lp R_{o}' \lp \frac{Z_{p}}{d} & j\omega L_{i}'\lp\frac{Z_{p}}{d} \\
        j\omega L_{i}' & j\omega L_{i}' \lp \frac{Z_{p}}{d} & j\omega L_{i}' \lp R_{i}' \lp \frac{Z_{p}}{d}
    \end{NiceArray}
    \right]
    \begin{bmatrix}
    J_{b} \\
    J_{o} \\
    J_{i}
    \end{bmatrix}.
\end{equation}
The schematic of the circuit corresponding to these coupled equations is shown in Fig. \ref{fig:SR_Short_Circuit}a. Note that the circuit is reciprocal, as it should be, which is reflected in the transpose symmetry of the impedance matrix above.

The above system of equations (\ref{eqn:ShortSR_Zmatrix}) is in block matrix form, just like (\ref{eqn:Block_Zmatrix_3x3}) and can by solved by the Schur complement (\ref{eqn:Schur_Soln_3x3_General}). After some algebra, the exact solution to the input impedance can be written as: 
\begin{equation*}
\begin{split}
    & Z_{in}=j\omega L_{c}- \\ 
    & \frac{(j\omega L_{i}')^{2}\left( 1\lp\left( \frac{L_{o}'}{L_{i}'} \right)^{2} \frac{R_{i}'}{R_{o}'} \lp \frac{Z_{p}(L_{o}'-L_{i}')^{2}}{dR_{0}L_{i}'^{2}} \lp \frac{j\omega(L_{o}'-L_{i}')}{R_{o}'} \frac{L_{o}'}{L_{i}'} \right)}{\frac{j\omega(L_{o}'-L_{i}')}{R_{o}'}\left( \frac{Z_{p}}{d} \lp j\omega L_{i}'+R_{i}' \right) \lp \left( \frac{Z_{p}}{d}\lp j\omega L_{i}' \lp \frac{R_{i}'R_{o}'}{R_{i}'\lp R_{o}'} \right)\left( 1\lp\frac{R_{i}'}{R_{o}'} \right) }.
\end{split}
\end{equation*}

\subsection{Approximate Permeability Assuming Thin Conductors}
The complexity of the preceding equation for the impedance $Z_{in}$ can be simplified significantly if we assume that the cross-sectional area of the conductor is a small fraction of the unit cell and that the conductor is much thicker than the skin depth, which imply the respective conditions: 
\begin{equation}
    \label{eqn:ShortSR_Thin_Assumptions}
    \frac{L_{o}'-L_{i}'}{L_{c}}\ll1, \quad \frac{R_{o,i}}{\omega(L_{o}'-L_{i}')} \edit{ \approx \frac{\delta}{t} \frac{2\pi a_{o,i}}{d} } \ll1.
\end{equation}
Given these assumptions, the input impedance for the Swiss roll unit cell can be approximated as: 
\begin{equation}
    \label{eqn:ShortSR_Zin_Approx}
    \boxed{ Z_{in}\approx j\omega L_{c}-\frac{ ( j\omega L_{i}')^{2} }{ \frac{1}{j\omega C'd} + j\omega L_{i}'+R_{i}' }. } 
\end{equation}

Before we give the associated effective permeability, note that the above input impedance has a common and recognizable form: it is the input impedance of an inductor $L_{c}$ that is inductively coupled to a series RLC resonator (Fig. \ref{fig:SR_Short_Circuit}c). The capacitor in the series RLC resonator is provided by the parallel plate capacitance of the waveguide, $C=\frac{\epsilon d}{g}$, while the inductance of the resonator $L_{i}'=L_{i}+\frac{L'd}{2}$ is due to the inductance of the current flowing around the interior of the Swiss roll, plus half of the total inductance of the parallel-plate waveguide. Why does only half of the waveguide inductance appear in the equation? An implication of the approximations made in finding (\ref{eqn:ShortSR_Zin_Approx}) is that no current flows on the outside of the Swiss roll, i.e., $J_o=0$. \edit{If} the current at the outer end of the waveguide must be zero, \edit{the current distribution in the waveguide is approximately triangular, decreasing linearly from the inner end of the guide to the outer end.} The total flux through the waveguide is \edit{therefore} half \edit{$L'd$, the total waveguide flux assuming} \edit{\sout{what it would be if}} the current \edit{\sout{in the waveguide}} is uniformly distributed\edit{\sout{ (which is the meaning of the total waveguide inductance $L'd$)}}. 

That no current flows on the exterior of the Swiss roll is a good approximation, except at low frequencies. Recognizing this allows us to map the approximation directly onto the circuit model in Fig. \ref{fig:SR_Short_Circuit}a -- just replace the inductor $L_o$ with an open circuit. Doing so results in the simplified circuit models in Fig. \ref{fig:SR_Short_Circuit}b-c, whose input impedances are given by (\ref{eqn:ShortSR_Zin_Approx}).

The associated effective permeability is, from (\ref{eqn:MuEff_Definition}), given by:
\begin{equation}
    \label{eqn:ShortSR_muEff_Apprx}
    \begin{split}
        \mu_{r} & =1-\frac{ j\omega\frac{L_{i}'^{2}}{L_{c}} }{j\omega L_{i}'+R_{i}'+\frac{1}{j\omega C}} \\
                & \approx 1-\frac{ \left( \frac{\pi a^{2}_{i}}{w_{c}^{2}} \right)\omega^{2} }{\omega^{2} -j\omega\frac{2\eta_{s}\mathcal{N}}{\mu_{0}a_{i}}-\frac{gc^{2}}{2\pi^{2} a^{3}_{i}\epsilon_{r}(\mathcal{N}-1)}}, 
    \end{split}
\end{equation}
where the expression in terms of geometric parameters has used the approximations: 
\begin{equation}
    \label{eqn:ShortSR_Circ_Params_Apprx}
    \begin{split}
        L_{i}' & =\mu_{o}A_{i}+\frac{gd}{2}\approx \mu_{0}\pi a_{i}^{2}\\
        C & =\frac{\epsilon d}{g}\approx \frac{\epsilon \left( 2\pi a_{i} \right)(\mathcal{N}-1)}{g} \\
        R'_{i} & \approx  \eta_{s}(2\pi a_{i})\mathcal{N}.
    \end{split} 
\end{equation}

The equations (\ref{eqn:ShortSR_muEff_Apprx})-(\ref{eqn:ShortSR_Circ_Params_Apprx}) are written to permit easy comparison to the permeability found in \cite{1999pendry_SwissRoll,2009demetriadou_SwissRollSims}, which are largely in agreement with our model. \edit{\sout{Aside from one quantitative difference in predicting the loss of the Swiss roll metamaterial, the power of our model for the Swiss roll is in the insight it gives.}} \edit{However, our circuit model} clarifies an interesting lacuna in the heuristic derivation of \cite{1999pendry_SwissRoll}. In that original work, inductance of the Swiss roll is thought to be proportional to the number of turns squared, as in a multi-coil inductor. Also, the capacitance of the Swiss roll is thought to be the series combination of $(\mathcal{N}-1)$ capacitors, whose total capacitance is \textit{inversely proportional} to the number of turns. In retrospect, this heuristic argument only predicts the correct resonant frequency by a felicitous accident. The natural resonant frequency $\omega_{0}$ of the permeability is $\omega_{0}=\frac{1}{\sqrt{ L_{i}'C }}$. The product $L_{i}'C$ depends linearly on the number of turns $(\mathcal{N}-1)$, since the inductance is (approximately) independent of the number of turns, while the capacitance increases linearly with the number of turns. However, if one calculates the resonant frequency thinking that the inductance is proportional to $(\mathcal{N}-1)^{2}$ while the capacitance is inversely proportional to $(\mathcal{N}-1)$, as in \cite{1999pendry_SwissRoll}, then $L_i'C$ is still linearly dependent on the number of turns, yielding the correct dependence of the resonant frequency on the number of turns in the Swiss roll. \edit{(In fact, the inductance of the Swiss roll \textit{does} increase with the number of turns, since the waveguide contributes to the inductance, a fact predicted by our model.)}

\edit{\sout{However, }}The heuristic derivation for the permeability of the Swiss roll \edit{also} breaks down in predicting its loss. While our model shows that the total loss of the metamaterial increases linearly with the number of turns of the Swiss roll, \cite{2009demetriadou_SwissRollSims} predicts that the loss actually is \textit{smaller} for larger number of turns. How are we to understand this discrepancy? It seems that the authors of \cite{2009demetriadou_SwissRollSims} viewed the current on adjacent turns of the Swiss roll as flowing parallel to each other, so that that if each full turn of the Swiss roll had a resistance of $R$, then the parallel combination of $(\mathcal{N}-1)$ such resistors had a total resistance of $\frac{R}{\mathcal{N}-1}$. This is inconsistent with the way current actually flows on the Swiss roll, passing sequentially through each turn of the Swiss roll -- so that the total resistance is the sum of the resistance of each turn, as indicated in (\ref{eqn:ShortSR_Circ_Params_Apprx}).

\subsection{The Lossless Swiss Roll}
The foregoing equation for the effective permeability (\ref{eqn:ShortSR_muEff_Apprx}) arose from approximating the conductor as negligibly thin compared to the rest of the unit cell, though still thick compared to the skin depth. If this approximation is not made, but the conductor is lossless, we obtain the \edit{\sout{(exact)}} input impedance:
\begin{equation}
    \label{eqn:ShortSR_Zin_Lossless}
    Z_{in}=j\omega L_{c}-\frac{(j\omega L_{i}')^{2}\left( \frac{Z_{p}}{dj\omega L_{i}'}  \frac{L_{o}'-L_{i}'}{L_{i}'}+  \frac{L_{o}'}{L_{i}'} \right)}{ \left( \frac{Z_{p}}{d} + j\omega L_{i}' \right)  }.
\end{equation}
In the low-frequency limit, the impedance of the lossless Swiss roll unit cell approaches: 
\begin{equation}
    \label{eqn:ShortSR_Limit_Zin_Lossless}
    \lim_{ \omega \to 0 }Z_{in}=j\omega(L_{c}-(L_{o}'-L_{i}')),
\end{equation}
with the corresponding effective permeability being
\begin{equation}
    \label{eqn:ShortSR_Limit_muEff_Lossless}
    \lim_{ \omega \to 0 } \mu_{zz} = 1 - \frac{A_{o}-A_{i}-gd}{w_{c}^{2}} = 1-\frac{A_{cond}}{w_{c}^{2}},  
\end{equation}
where $A_{cond}$ is the cross-sectional area of the unit cell that is taken up by the perfect conductor. This makes sense, as no flux may pass through a perfect conductor; the relative permeability of unit cells containing circular cylinders was reduced from unity by the fill-fraction for the same reason in Section \ref{sec:Circular_Cylinders}. 

Having verified that the Swiss roll model predicts behavior that is consistent with our intuitions, \sout{and having clarified the ways that the model differs from that of ,} we are now ready to \edit{\sout{move on to generalizing}}\edit{generalize} the model to include higher order modes.

\section{Swiss Roll Metamaterial with a Spiral of Arbitrary Length}
\label{sec:SR_AnyLength}
In the prior section, we analyzed the Swiss roll unit cell for cases where the length of the \edit{\sout{PPW}} \edit{spiral waveguide} was electrically short $(d\ll \lambda)$. Now, let us turn our attention to the general case, when \edit{\sout{the PPW} $d$} can be any length. \edit{\sout{This problem seems deceptively simple, since the parallel-plate waveguide is a well-known structure, and we'll only be exciting the TEM mode of the guide. However, this is no ordinary PPW. Rather, it is a flux-coupled waveguide, where}} \edit{As before,} every small section of the spiral waveguide is inductively coupled to currents flowing outside of the guide. The analytic solution of the problem (which is reminiscent of the method of moments) yields the higher order modes of the Swiss roll metamaterial\edit{\sout{, and also illuminates the solution to problems in electromagnetics in which symmetric nearest-neighbor coupling is present}}. 

\subsection{Problem Setup}
\label{subsec:LongSR_Setup}
As in Section \ref{sec:ShortSR}, the Swiss roll unit cell is divided into subregions, where a uniform current flows around the boundary of each region and produces a uniform axial magnetic field interior to that region. Three of these subregions are already familiar from our prior analysis: one which is defined by the boundary of the unit cell $\mathcal{C}_{b}$, and two defined by the inner $\mathcal{C}_{i}$ and outer $\mathcal{C}_{o}$ perimeters of the Swiss roll. For the naming of the associated magnetic fields, electric currents, and inductances, see Table \ref{table:SubregionProperties}. In addition, the PPW is divided into $N$ subregions of identical length, $\Delta \coloneqq \frac{d}{N}$ and height $g$ (the \edit{size of the gap between the conductors}). Each of these regions is bounded by a closed curve $\mathcal{C}_{k}$ on which current $J_{k}$ flows, producing a uniform axial magnetic field $H_{z}^{k}=J_{k}$; \edit{\sout{where}} $k\in \{ 1,2,\dots,N \}$ indexes over the subregions of the \edit{\sout{PPW} guide}.

\begin{figure*}
\centering
\includegraphics[width=2.00\columnwidth]{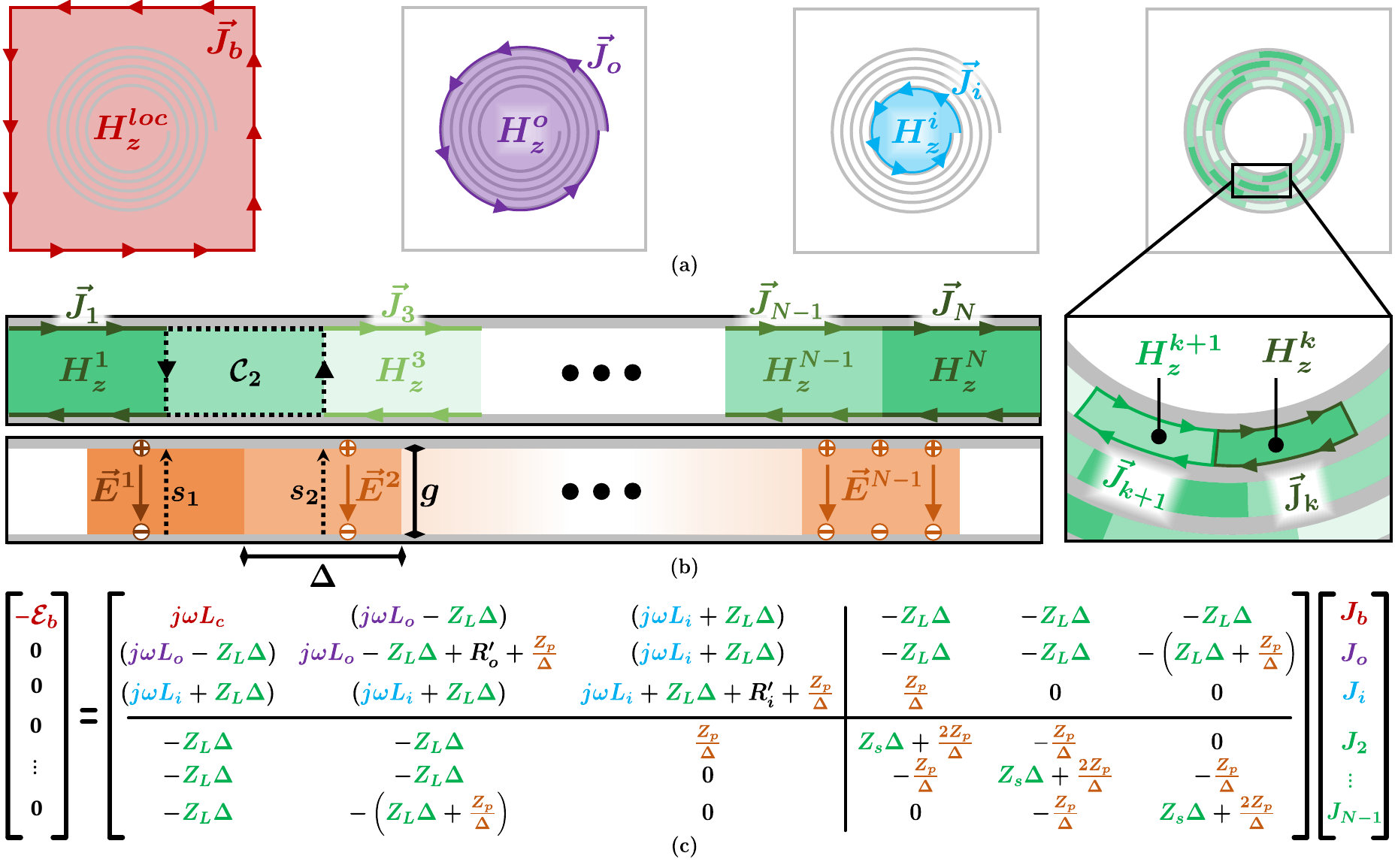}
\caption{Problem setup for the Swiss roll unit cell for arbitrary \edit{length} $d$ \edit{of the spiral waveguide}. (a) The unit cell is divided into subregions (shaded), and a uniform current distribution is assumed to flow around the perimeter of each subregion. Each current distribution produces a uniform axial magnetic field within its subregion. (b) The spiral of the Swiss roll is broken up into $N$ subregions. Straightening the spiral out turns these into sections of PPW of total length $d$. If the currents in adjacent sections are not equal, then charge accumulates on the top and  bottom conductors, producing an electric field. This electric field is discretized along a grid that is dual to that for the magnetic field. (c) The large matrix equation resulting from repeated application of Faraday's law in integral form, with colors indicating the subregions from which the different impedances arise.}
\label{fig:Long_SwissRoll}
\end{figure*}

The electric field within the \edit{\sout{PPW} gap} is similarly discretized into piecewise-uniform subregions. These subregions are also of length $\Delta$, but are discretized over a grid that is dual to the magnetic field subregions, as shown in Fig. \ref{fig:Long_SwissRoll}. Discretizing the electric and magnetic fields over dual grids is necessary for preserving the self-dual structure of Maxwell's equations in homogeneous media, as in the Yee grid for finite-difference methods \cite{1966_yeeGrid,2019nakata_SelfDuality}.

\edit{\sout{Of the different subregions that have been defined, }}Let us begin by analyzing the $k^{th}$ subregion of the \edit{\sout{PPW} guide}, where $k \in \{ 2,\dots,N-1 \}$ so that the subregion is interior to the guide, not on either of its ends. Application of Faraday's law (\ref{eqn:FaradayGeneral}) around $\mathcal{C}_{k}$, accounting for mutual flux from the currents $J_{o}$ and $J_{b}$, gives:
\begin{equation}
    \label{eqn:LongSR_PPW_Faraday}
    -\mathcal{E}_{k-1}^{s}+\mathcal{E}_{k}^{s}=-Z_{L}\Delta(J_{b}+J_{o})+Z_{s}\Delta J_{k}.
\end{equation}
Expressing the EMF $\mathcal{E}_{k}$ in terms of electric currents requires the application of Gauss's law. The line integral for $\mathcal{E}_{k}^{s}$ is taken at the boundary between regions $k$ and $k+1$ of the waveguide, which carry currents $J_{k}$ and $J_{k+1}$, respectively. By conservation of charge, a charge equal to 
\begin{equation}
    \label{eqn:LongSR_Qk}
    Q_{k}=\frac{1}{j\omega}\left( J_{k}-J_{k+1} \right)
\end{equation}
must accumulate over this boundary. Assuming that this charge is uniformly distributed within the corresponding constant-$E$ subregion, application of Gauss's law yields the electric field and the associated EMF:
\begin{equation}
    \label{eqn:LongSR_Ek_ito_Jk}
    \mathcal{E}_{k}^{s}=-\frac{Z_{p}}{\Delta}(J_{k}-J_{k+1}).
\end{equation}
Substituting this equation into (\ref{eqn:LongSR_PPW_Faraday}) allows us to express the behavior of the $k^{th}$ \edit{\sout{PPW} waveguide} section entirely in terms of electric currents: 
\begin{equation}
    \label{eqn:Long_SR_KVL_PPW_Section}
    \begin{split}
        0 & =-Z_{L}\Delta \left( J_{b}+J_{o} \right)-\frac{Z_{p}}{\Delta}J_{k-1}\\
        & +\left( Z_{s}\Delta+\frac{2Z_{p}}{\Delta} \right)J_{k}-\frac{Z_{p}}{\Delta}J_{k+1}.
    \end{split}
\end{equation}
A similar process yields the circuit equations for the first and final segments of the \edit{\sout{PPW} waveguide}: 
\begin{equation}
    \label{eqn:LongSR_KVL_PPW_Ends}
    \begin{split}
        \lm \mathcal{E}_{0}^s = & - Z_{L}\Delta(J_{b}+J_{o})+\left( Z_{s}\Delta+\frac{Z_{p}}{\Delta} \right)J_{1}-\frac{Z_{p}}{\Delta}J_{2} \\
        \mathcal{E}_{N}^s = & - Z_{L}\Delta(J_{b}+J_{o}) - \frac{Z_{p}}{\Delta}J_{N-1} \\
        +&\left( Z_{s}\Delta+\frac{Z_{p}}{\Delta} \right)J_{N},
    \end{split}
\end{equation}
Now that we have found the circuit equations for the loops $\mathcal{C}_{k}$\edit{\sout{in the PPW}}, we can complete the setup by writing circuit equations for the three remaining loops $\mathcal{C}_{b}$, $\mathcal{C}_{o}$, and $\mathcal{C}_{i}$. Applying Faraday's law and taking care to account for the mutual flux induced into $\mathcal{C}_{b}$ and $\mathcal{C}_{o}$ by the currents in the \edit{\sout{PPW} waveguide}, these equations are\edit{\sout{ given by}}: 
\begin{align}
\label{eqn:LongSR_Faraday_3OrdinaryLoops}
    -\mathcal{E}_{b} & = j\omega L_{c}J_{b}+j\omega L_{o}J_{o}+j\omega L_{i}J_{i}- \sum_{k=1}^{N}(Z_{L}\Delta)J_{k} \nonumber \\
    -\mathcal{E}_{N}  & = j\omega L_{o}J_{b}+(j\omega L_{o}+R_{o})J_{o}+j\omega L_{i}J_{i}- \sum_{k=1}^{N}(Z_{L}\Delta)J_{k} \nonumber \\ 
    -\mathcal{E}_{0} & = j\omega L_{i}J_{b}+j\omega L_{i}J_{o}+(j\omega L_{i}+R_{i})J_{i},
\end{align}
where $\mathcal{E}_{b}$ is the EMF around the boundary of the cell \edit{\sout{(\ref{eqn:Ann_Cyl_EMF_integral})}} and the inductances are as defined in Table \ref{table:SubregionProperties}. The resistances $R_{o}$ and $R_{i}$ are proportional to the lengths of the conductors adjacent to the $\mathcal{C}_{o}$ and $\mathcal{C}_{i}$, respectively: 
\begin{equation}
    \label{eqn:LongSR_Ro_Ri_Def}
    R_{o,i}=\eta_{s}\ell_{o,i}.
\end{equation}
Current continuity at the two ends of the PPW requires that
\begin{equation}
    \label{eqn:LongSR_KCL}
    J_{1}=-J_{i}, \quad J_{N}=J_{o}.
\end{equation}
Enforcing continuity of current, then substituting (\ref{eqn:LongSR_KVL_PPW_Ends}) into (\ref{eqn:LongSR_Faraday_3OrdinaryLoops}) yields a system of $(N+1)$ linear equations. Written in block-matrix form, the system of equations is:
\begin{equation}
\label{eqn:LongSR_Block_Zmat_Full}
\begin{bmatrix}
    \mathbf{V}_{b} \\
    0
    \end{bmatrix}
    =
    \begin{bmatrix}
    \mathbf{Z}_{11} & \mathbf{Z}_{12} \\
    \mathbf{Z}_{21} & \mathbf{Z}_{22}
    \end{bmatrix}
    \begin{bmatrix}
    \mathbf{I}_{1} \\
    \mathbf{I}_{2}
\end{bmatrix}.
\end{equation}
The full system of equations is shown in Fig. \ref{fig:Long_SwissRoll}c, and the primed resistances from the figure are defined by: 
\begin{equation}
    \label{eqn:LongSR_Rprime_Def}
    R_{o,i}'\coloneqq R_{o,i}+R'\Delta.
\end{equation}
Note that the full impedance matrix is transpose symmetric, since the system is reciprocal. \edit{\sout{Additionally, the submatrix $\mathbf{Z}_{11}$ is identical in form to the matrix in (\ref{eqn:ShortSR_Zmatrix}), which is the system of equations obtained assuming $d\ll \lambda$. This makes sense, as submatrix $\mathbf{Z}_{11}$ embeds information about the cell subregion, the inner and outer subregions, and the two waveguide subregions closest to the ends of the Swiss roll, which are the same subregions are used in setting up the problem in Section \ref{sec:ShortSR}.}}

The formal solution to (\ref{eqn:LongSR_Block_Zmat_Full}) is given by the Schur complement: 
\begin{equation}
    \label{eqn:LongSR_Block_Schur_Formal_Soln}
    \mathbf{V}_{b}=(\mathbf{Z}_{11}-\mathbf{Z}_{12}\mathbf{Z}_{22}^{-1}\mathbf{Z}_{21})\mathbf{I}_{1}.
\end{equation}
Analytically solving this system of equations is made possible by examining the eigenvectors and eigenvalues of the submatrix $\mathbf{Z}_{22}$, the impedance matrix associated with the electric current distribution inside the PPW. 

\subsection{Solving the Schur Complement}
\label{subsec:LongSR_SchurComplement}
No matter how many subsections are used to discretize the field inside of the PPW, the submatrix $\mathbf{Z}_{22}$ always takes the form of a symmetric, tridiagonal Toeplitz matrix of size $(M-1)\times (M-1)$ with $M\coloneqq N-1$: 
\begin{equation}
    \label{eqn:Toeplitz_Matrix}
    \mathbf{Z}_{22}=\begin{bmatrix}
    a & b & 0 & 0 & \cdots &  0 \\
    b & a & b & 0 &  & 0 \\
    0 & b & a & b &  & 0 \\
    0 & 0 & b & a  &  & 0 \\
    \vdots &  &  &  & \ddots &  \vdots \\
    0 & 0 & 0 & 0 & \cdots &  a
    \end{bmatrix},
\end{equation}
where \edit{$a=\left( Z_{s}\Delta+\frac{2Z_{p}}{\Delta} \right)$ and $b=-\frac{Z_{p}}{\Delta}$}. The matrix is Toeplitz due to the (discrete) shift-invariance of the model, i.e., the same equation (\ref{eqn:Long_SR_KVL_PPW_Section}) applies to every subregion of the PPW. Because the loop equation for section $k$ of the PPW (\ref{eqn:Long_SR_KVL_PPW_Section}) depends on the current in that section and upon its nearest neighbors, the matrix is tridiagonal. Transpose symmetry arises due to reciprocity.

As shown in Appendix \ref{Appendix_Toeplitz}, the submatrix $\mathbf{Z}_{22}$ can be unitarily diagonalized as: 
\begin{equation}
    \label{eqn:Z22_Diagonalized}
    \mathbf{Z}_{22}=\mathbf{U}\mathbf{D}\mathbf{U},
\end{equation}
where $\mathbf{U}=\begin{bmatrix} \hat{\mathbf{I}}_{1} & \hat{\mathbf{I}}_{2} & \cdots & \hat{\mathbf{I}}_{N-2}\end{bmatrix}$ is the matrix of eigenvectors, which is unitary and transpose symmetric, and $\mathbf{D}$ is the diagonal matrix of eigenvalues: 
\begin{equation}
    \label{eqn:Diagonal_Matrix}
    \mathbf{D}_{\ell,m}=\
        \begin{cases}
        \frac{Z_{p}}{\Delta}\left( \frac{Z_{s}}{Z_{p}}\Delta^{2}+4\sin^{2}\left( \frac{m\pi}{2M} \right) \right)  & \ell=m \\
        0  & \ell\neq m
        \end{cases}.
\end{equation}
The eigenvectors of the matrix $\mathbf{Z}_{22}$ can be physically interpreted as modal current distributions within the parallel-plate waveguide, where the current at the ends of the guide is required to be zero. 

It will be useful to define the vector $\mathbf{V}$ that is the sum of all $(N-2)$ eigenvectors. This vector is given by:
\begin{equation}
    \label{eqn:Sum_of_Eigenvectors}
    [\mathbf{V}]_{k}=\sqrt{ \frac{2}{M} } \sin^{2}\left( \frac{k\pi}{2} \right) \cot\left( \frac{k\pi}{2M} \right),
\end{equation}
where the only nonzero entries of this vector are those for which the index $k$ is odd.

We now endeavor to directly solve the Schur complement (\ref{eqn:LongSR_Block_Schur_Formal_Soln}). To do this, note that because $\mathbf{Z}_{22}$ can be unitarily diagonalized as in (\ref{eqn:Z22_Diagonalized}), its inverse can be written: 
\begin{equation}
    \label{eqn:LongSR_Z22_Inv}
    \mathbf{Z}_{22}^{-1}=\mathbf{U}\mathbf{D}^{-1}\mathbf{U}.  
\end{equation}
The matrix $\mathbf{Z}_{21}$ takes a simple form: 
\begin{equation}
    \label{eqn:LongSR_Z21}
    \mathbf{Z}_{21}
    =
    \begin{bmatrix}
    -(Z_{L}\Delta) \mathbf{1} & -(Z_{L}\Delta)\mathbf{1}-\frac{Z_{p}}{\Delta}\hat{\mathbf{e}}_{N-2} & \frac{Z_{p}}{\Delta}\hat{\mathbf{e}}_{1}
    \end{bmatrix},
\end{equation}
where $\hat{\mathbf{e}}_{k}$ are standard Cartesian unit vectors of size for a vector space of $(N-2)$ dimensions and $\mathbf{1}$ is the sum of the standard Cartesian unit vectors, i.e., it is a vector whose rows are all equal to 1. Consequently, the matrix product $\mathbf{Z}_{12}\mathbf{Z}_{22}^{-1}\mathbf{Z}_{12}$ within the Schur complement can be written as: 
\begin{equation*}
    \label{eqn:LongSR_Zinv}
    \begin{bmatrix}
    Z_{L}^{2}A & Z_{L}^{2}A \lp Z_{L}Z_{p}C & \lm Z_{L}Z_{p}B \\
    Z_{L}^{2}A  \lp  Z_{L}Z_{p}C & Z_{L}^{2}A \lp 2Z_{L}Z_{p}C \lp Z_{p}^{2}F & \lm Z_{L}Z_{p}B\lm Z_{p}^{2}E \\
    \lm Z_{L}Z_{p}B & \lm Z_{L}Z_{p}B\lm Z_{p}^{2}E  & Z_{p}^{2}D
\end{bmatrix}
\end{equation*}
where the scalars $A-F$ are defined by the following products:
\begin{align}
A  & =\Delta^{2}\mathbf{V}^T\mathbf{D}^{-1}\mathbf{V} \\
B  & =\hat{\mathbf{I}}_{1}^{T}\mathbf{D}^{-1}\mathbf{V} \\
C  & =\hat{\mathbf{I}}_{N-2}^{T}\mathbf{D}^{-1}\mathbf{V} \\
D  & = \frac{1}{\Delta^{2}}\hat{\mathbf{I}}_{1}^{T}\mathbf{D}^{-1}\hat{\mathbf{I}}_{1}  \\
E  & = \frac{1}{\Delta^{2}}\hat{\mathbf{I}}_{N-2}^{T}\mathbf{D}^{-1}\hat{\mathbf{I}}_{1} \\
F  & = \frac{1}{\Delta^{2}}\hat{\mathbf{I}}_{N-2}^{T}\mathbf{D}^{-1}\hat{\mathbf{I}}_{N-2}.
\end{align}

It is possible to show that $B=C$ and $D=F$, so that there are only four scalar quantities that must be computed in order to fully populate the matrix. Each of these quantities can be written as sums, and each of these sums has a well-defined limit \edit{\sout{for large}}\edit{as the number of waveguide sections} $N$ \edit{goes to infinity}. \edit{\sout{Finding these equations involves approximating the terms in the sum when $M=N-1$ is large, rearranging the sum until it takes a recognizable form, and then allowing the sum to contain an infinite number of terms, i.e., taking the limit as $N \rightarrow \infty$.}}  \textbf{To simplify the notation involved, we use the symbol $\asymp$ to denote asymptotic equivalence in the limit of large $N$.}

Before solving the sums, let us define the quantity $\gamma$,
\begin{equation}
    \label{eqn:Gamma_Def}
    \gamma \coloneqq -j\sqrt{ \frac{Z_{s}}{Z_{p}} }d,
\end{equation}
which will appear repeatedly. The physical meaning of this quantity \edit{\sout{becomes} is} clear if we consider \edit{\sout{what it becomes}its value} when the conductor is lossless. Using the definitions of the impedances \edit{\sout{and their associated per-unit-length quantities}} (\ref{eqn:PPW_Zs_Zp_Defs}), we find that it is simply the propagation constant multiplied by the length of the waveguide:
\begin{equation}
    \label{eqn:Gamma_eq_kd}
    \gamma=-j\sqrt{ j\omega L'\cdot j\omega C' }d=\omega\sqrt{ \mu_{0}\epsilon } d=kd.
\end{equation}

Written in sum form, the scalar A is equal to: 
\begin{equation}
\label{eqn:LongSR_A_sum}
    A= \frac{2\Delta^{2}}{M} \sum_{k=1}^{M-1}\frac{\sin^{4}\left( \frac{k\pi}{2} \right)\cot^{2}\left( \frac{k\pi}{2M} \right)}{\frac{Z_{p}}{\Delta}\left( \frac{Z_{s}}{Z_{p}}\Delta^{2}+4\sin^{2}\left( \frac{k\pi}{2M} \right) \right)}.
\end{equation}
This sum is really over positive, odd indices, since all of the even terms are zero; we denote this set by $\mathbb{N}_{odd}=\{ 1,3,5,\dots \}$. Approximating the summand when the number of terms in the sum is allowed to be arbitrarily large gives: 
\begin{equation}
\label{eqn:LongSR_A_approx}
    \begin{split}
        A & \asymp \frac{2d^{3}}{Z_{p}M^{4}} \sum_{k\in \mathbb{N}_{odd}} \frac{\left( \frac{2M}{k\pi} \right)^{2}}{\left( \frac{k\pi}{M} \right)^{2}-\frac{\gamma^{2}}{M^{2}}} \\
         & = \frac{d^{2}}{Z_{p}/d} \sum_{k\in\mathbb{N}_{odd}} \frac{8}{(k\pi)^{2}\left( (k\pi)^{2}-\gamma^{2} \right)} \\
         & =\frac{8d^{2}}{Z_{p}/d} \frac{1}{\gamma^{2}} \sum_{k\in \mathbb{N}_{odd}} \left( \frac{1}{(k\pi)^{2}-\gamma^{2}} - \frac{1}{(k\pi)^{2}} \right).
    \end{split}
\end{equation}
The final term in the summand is the reciprocal-squares of the odd integers, and its sum can be found using the solution to the famous Basel problem. The first term in the summand is closely related to the expansion of the tangent function in terms of its poles by the Mittag-Leffler theorem: 
\begin{equation}
    \tan\left( \frac{x}{2} \right)=4 \sum_{k\in\mathbb{N}_{odd}} \frac{x}{(k\pi)^{2}-x^{2}}.
\end{equation}
Therefore, the limit of $A$ as the number of subsections in the PPW goes to infinity is given by: 
\begin{equation}
\label{eqn:LongSR_A}
    A\asymp \frac{d^{2}}{Z_{s}d}\left[ 1-\frac{\tan(\gamma/2)}{(\gamma/2)} \right].
\end{equation}

The process is simpler for the quantity $B$: 
\begin{equation}
\label{eqn:LongSR_B}
    \begin{split}
        B & = \frac{2}{M} \sum_{k=1}^{M-1} \frac{\sin^{2}\left( \frac{k\pi}{2} \right)\sin\left( \frac{k\pi}{M} \right)\cot\left( \frac{k\pi}{2M} \right)}{\frac{Z_{p}}{\Delta}\left( \frac{Z_{s}}{Z_{p}}\Delta^{2}+4\sin^{2}\left( \frac{k\pi}{2M} \right) \right)} \\
          & \asymp \frac{2d}{Z_{p}M^{2}} \sum_{k\in \mathbb{N}_{odd}} \frac{\left( \frac{k\pi}{M} \right)\left( \frac{2M}{k\pi} \right)}{\left( \frac{k\pi}{M} \right)^{2}-\frac{\gamma^{2}}{M^{2}}} \\
          & = \frac{1}{Z_{p}/d } \sum_{k\in \mathbb{N}_{odd}} \frac{4}{(k\pi)^{2}-\gamma^{2}} \\
          & = \frac{1}{Z_{p}/d } \frac{\tan(\gamma/2)}{\gamma}.
    \end{split}
\end{equation}

\begin{figure}[!t]
\centering
\includegraphics[width=1\columnwidth]{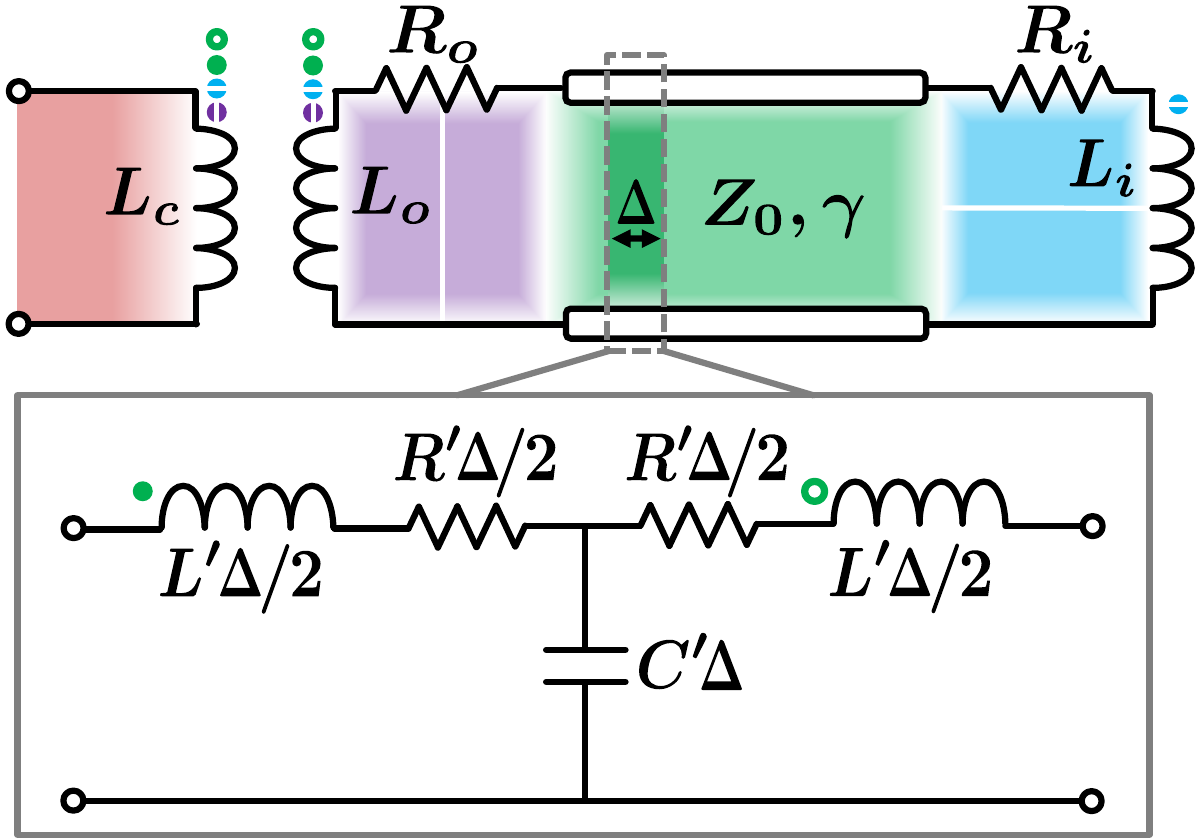}
\caption{The distributed circuit model of the Swiss roll unit cell. The inductances $L_{o,i}$ are associated with currents flowing on the outside and inside surfaces of the Swiss roll, $L_{c}$ is the inductance of an empty unit cell, and the transmission line corresponds to the spiral waveguide formed by the Swiss roll. This is a ``flux coupled" transmission line, since each differential section of the line is magnetically coupled to other portions of the unit cell. A circuit model for a differential section (width $\Delta$) of the flux-coupled transmission line is shown at the bottom. Primed quantities are per-unit-length parameters. The system of equations corresponding to the circuit model is (\ref{eqn:Double_Col}).}
\label{fig:LongSR_Circuit}
\end{figure}

Solving for D, however, requires extra caution: there is a term in the sum that grows (linearly) as the number of divisions of the waveguide is increased. To address this, we sequester this term and ignore the fact that it has no limit. Our patience will be rewarded when this term exactly cancels the only entry in $\mathbf{Z}_{11}$ that also ``blows up" as the number of divisions in the waveguide goes to infinity. We will also need to use the \edit{\sout{following}} expansion of the cotangent function in terms of its poles: 
\begin{equation}
    \label{eqn:LongSR_cotangent_series}
    \cot(x)=\frac{1}{x}-\sum_{k=1}^{\infty} \frac{2x}{(k\pi)^{2}-x^{2}}.
\end{equation}
Writing the explicit sum for $D$, we find:
\begin{equation}
    \label{eqn:LongSR_D}
    \begin{split}
        D & = \frac{2}{Z_{p}d } \sum_{k=1}^{M-1} \frac{\sin^{2}\left( \frac{k\pi}{M} \right)}{\frac{Z_{s}}{Z_{p}}\Delta^{2} + 4 \sin^{2}\left( \frac{k\pi}{2M} \right) } \\
          & \asymp \frac{1}{Z_{p}d}\sum_{k=1}^{M-1}\left( \frac{2\gamma^{2}}{(k\pi)^{2}-\gamma^{2}} + \cos\left( \frac{k\pi}{M} \right) +1   \right) \\
          & \asymp \frac{1}{Z_{p}d}[-\gamma \cot(\gamma)+(M+1)] \\
          & = \frac{1}{Z_{p}^{2}}\left( j\sqrt{ Z_{s}Z_{p} }\cot(\gamma) + \frac{Z_{p}}{\Delta} \right). 
    \end{split} 
\end{equation}
Because the differential width $\Delta$ approaches zero as the number of waveguide sections goes to infinity, it is the final term of this equation which has no limit for large $N$. 

Solving for the last unknown scalar, $E$, requires the expansion of the cosecant function in terms of its poles: 
\begin{equation}
    \label{eqn:LongSR_csc_expansion}
    \csc(x)=\frac{1}{x}+\sum_{k=1}^{\infty} \frac{(-1)^{k}(2x)}{x^{2}-(k\pi)^{2}},
\end{equation}
along with a certain series identity: 
\begin{equation}
    \sum_{k=1}^{M-1}(-1)^{k}\cos^{2}\left( \frac{k\pi}{2M} \right)=-\frac{1}{2}.
\end{equation}
The sum for $E$ yields, in the large-$N$ limit:
\begin{equation}
    \label{LongSR_E}
    \begin{split}
        E & = \frac{2}{Z_{p}d} \sum_{k=1}^{M-1}  \frac{\sin\left( \frac{k\pi}{M} \right)\sin\left( k\pi\frac{M-1}{M} \right)}{\frac{Z_{s}}{Z_{p}}\Delta^{2}+4\sin^{2}\left( \frac{k\pi}{2M} \right)} \\
          & \asymp \frac{2}{Z_{p}d}\sum_{k=1}^{M-1} \left( \frac{(-1)^{k}\gamma^{2}}{\gamma^{2}-(k\pi)^{2}} -(-1)^{k}\cos^{2}\left( \frac{k\pi}{2M} \right)\right) \\
          & \asymp -j \frac{\sqrt{ Z_{s}Z_{p} }}{Z_{p}^{2}} \csc(\gamma).
    \end{split}
\end{equation}
In order to make the following equation fit within the bounds of the page, we define the quantities: 
\begin{equation}
    \label{eqn:LongSR_Trig_Abbreviations}
    \begin{split}
        \tau  & \coloneqq \frac{\tan(\gamma/2)}{\gamma} \\
        \alpha & \coloneqq \cot(\gamma) \\
        \beta  & \coloneqq \csc(\gamma).
    \end{split} 
\end{equation}
The exact equation for the Schur complement $\mathbf{Z}_{S}=\mathbf{Z}_{11}-\mathbf{Z}_{12}\mathbf{Z}_{22}^{-2}\mathbf{Z}_{21}$ can now be written in the limit as the number of waveguide subdivisions goes to infinity (\ref{eqn:Double_Col}). 

\begin{strip}
\hrulefill 
    \begin{equation}
        \label{eqn:Double_Col}
        \begin{bmatrix}
            \lm \mathcal{E}_c \\ 0 \\ 0
        \end{bmatrix}
        =
        \begin{bmatrix}
        j\omega L_{c} \lm \frac{(Z_{L}d)^{2}}{Z_{s}d}(1 \lm 2\tau) & j\omega L_{o} \lm \frac{(Z_{L}d)^{2}}{Z_{s}d}(1 \lm 2\tau) \lm (Z_{L}d)\tau & j\omega L_{i} \lp (Z_{L}d)\tau \\
        j\omega L_{o} \lm \frac{(Z_{L}d)^{2}}{Z_{s}d}(1 \lm 2\tau) \lm (Z_{L}d)\tau & j\omega L_{o}\lp R_{o} \lm \frac{(Z_{L}d)^{2}}{Z_{s}d}(1 \lm 2\tau) \lm 2(Z_{L}d)\tau  \lm  j\sqrt{ Z_{s}Z_{p} }\alpha & j\omega L_{i}\lp (Z_{L}d)\tau \lm j\sqrt{ Z_{s}Z_{p} }\beta \\
        j\omega L_{i}\lp (Z_{L}d)\tau & j\omega L_{i}\lp (Z_{L}d)\tau \lm j\sqrt{ Z_{s}Z_{p} }\beta & j\omega L_{i}\lp R_{i}  \lm  j\sqrt{ Z_{s}Z_{p} }\alpha
        \end{bmatrix}
        \begin{bmatrix}
            J_b \\ J_o \\ J_i
        \end{bmatrix}
        .
    \end{equation}
\end{strip}

The circuit model corresponding to this system of equations is shown in Fig. \ref{fig:LongSR_Circuit}, which includes a flux-coupled transmission line. This flux-coupled transmission line accounts for the mutual coupling between each differential section of the transmission line \edit{\sout{(each region of constant magnetic field)}} and the rest of the unit cell. The form of the system of equations (\ref{eqn:Double_Col}) is the same as (\ref{eqn:Block_Zmatrix_3x3}), and can be solved by the Schur complement (\ref{eqn:Schur_Soln_3x3_General}). Letting $[\mathbf{Z}_{S}]_{m,n}=Z_{mn}$, the input impedance for the system of equations (\ref{eqn:Double_Col}) is formally\edit{\sout{given by}}: 
\begin{equation}
\label{eqn:LongSR_Zin_Schur_Zij}
    Z_{in}=Z_{11}\lm \frac{Z_{12}\left( Z_{21}Z_{33}\lm Z_{21}Z_{23} \right) \lp Z_{13}\left( Z_{31}Z_{22}\lm Z_{21}Z_{33} \right)}{Z_{22}Z_{33}\lm Z_{23}Z_{32}},
\end{equation}
which can be used to find the effective permeability of the cell according to (\ref{eqn:MuEff_Definition}).

\begin{figure*}[t]
    \centering
    \subfloat[ ]{\includegraphics[width=1.09\columnwidth]{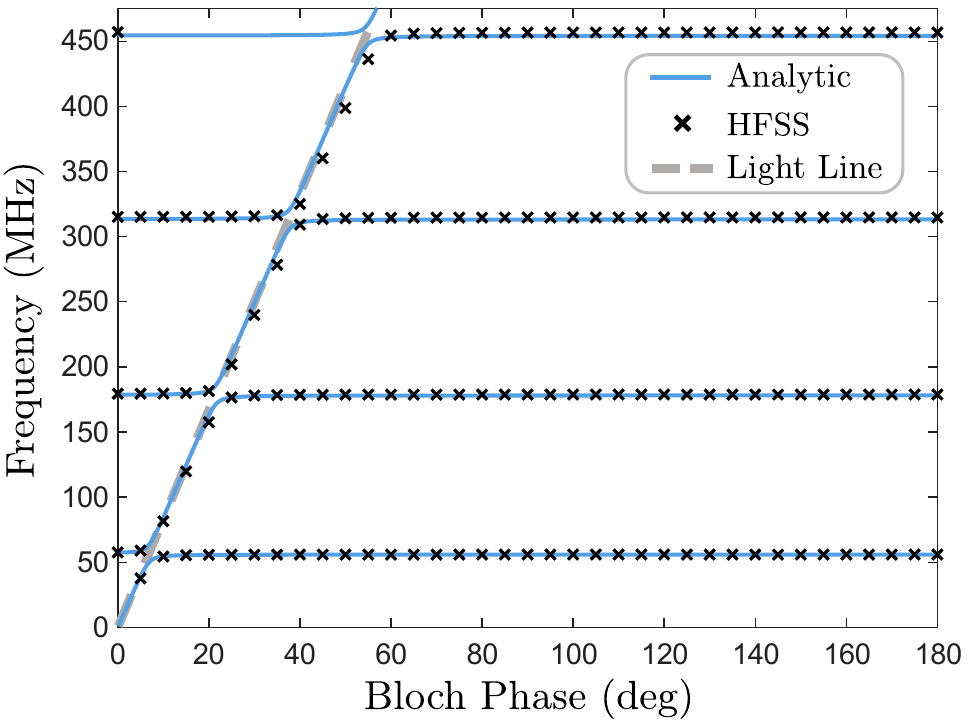}}%
    \hfil
    \subfloat[ ]{\includegraphics[width=0.90\columnwidth]{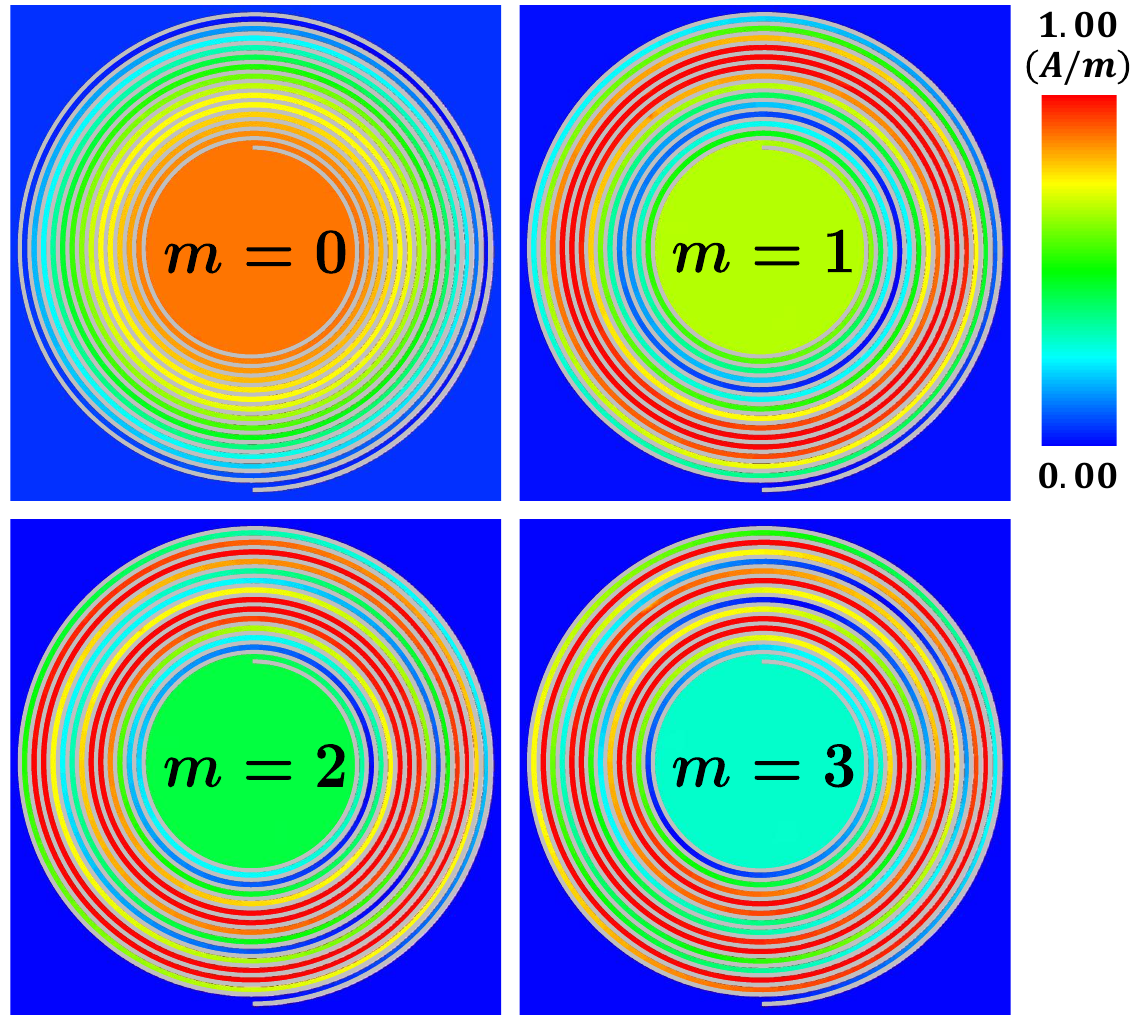}}
    \caption{Simulations of the Swiss roll metamaterial. (a) A comparison of the dispersion diagrams from the analytic model of the Swiss roll metamaterial and from full-wave simulation in Ansys HFSS. Here the unit cell is designed to be a tenth of a wavelength wide at 300 MHz and the spiral waveguide of the Swiss roll is a full wavelength long at the same frequency. (b) The axial magnetic field magnitude within the Swiss roll for each of the four modes when the Bloch phase is zero degrees. Note that as the mode index increases, \edit{\sout{the}} less magnetic field exists in the center of the Swiss roll compared to the spiral waveguide. The mode index corresponds to the number of field maxima (dark red) in the standing wave patterns inside the spiral. }
    \label{fig:Results}
\end{figure*}

Given the complexity of (\ref{eqn:LongSR_Zin_Schur_Zij}), it would be good to know if there is a simple approximate form of the input impedance in the low-loss limit. Sparing the reader some tedious algebra, the answer is yes. Specifically, if the loss-tangent associated with the propagation constant is very small, i.e., 
\begin{equation}
    \frac{\mathrm{Im}(\gamma)}{\mathrm{Re}(\gamma)}\ll1,
\end{equation}
the impedance of the Swiss roll unit cell \edit{is approximately}: 
\begin{equation}
\label{eqn:LongSR_Zin_Apprx}
    \boxed{
    \begin{aligned}
        Z_{in} & \approx j\omega (L_{c}-(L_o-L_i))+2jZ_{0}\tan\left( \frac{\gamma}{2} \right) \\
        & -\frac{\left( j\omega L_{i}+jZ_{0}\tan\left( \frac{\gamma}{2} \right) \right)^{2}}{j\omega L_{i}+R_{i}-jZ_{0}\cot(\gamma)},
    \end{aligned}
    }
\end{equation}
where $Z_{0}$ is the characteristic impedance of the \edit{\sout{parallel-plate}} waveguide: 
\begin{equation}
    Z_{0}=\sqrt{ Z_{s}Z_{p} }=\sqrt{ \frac{L'}{C'}-j\frac{R'}{\omega C'} }.
\end{equation}
Since (\ref{eqn:LongSR_Zin_Apprx}) was derived only assuming a low-loss conductor, the equation is exact in the lossless limit.

The input impedance (\ref{eqn:LongSR_Zin_Apprx}), like the input impedances (\ref{eqn:PEC_Cyl_Zin}) and (\ref{eqn:ShortSR_Zin_Approx}), takes the general form, 
\begin{equation}
    \label{eqn:Results_Z_Inversion}
    Z_{in}=Z_{11} - \frac{Z_c^2}{Z_L},
\end{equation}
which is the input impedance looking into an impedance inverter having coupling impedance $Z_c$ and \edit{load\sout{loaded with}} impedance $Z_L$. When the coupling impedance is inductive, as in (\ref{eqn:ShortSR_Zin_Approx}), this simply indicates that the impedance inversion arises from transforming the load impedance through a mutual inductance (as in Fig. \ref{fig:SR_Short_Circuit}c). In the case of (\ref{eqn:LongSR_Zin_Apprx}), the coupling impedance is
\begin{equation}
    Z_c = j\omega L_i+jZ_0\tan(\gamma/2).
\end{equation}
The fact that the coupling impedance includes the tangent is explained by the flux-coupling between the waveguide and the unit cell. That is, any flux inside the spiral of the Swiss roll is mutually coupled to the inductance $L_c$ of the entire unit cell. It is normally somewhat strange for a distributed circuit like the flux-coupled transmission line in Fig. \ref{fig:LongSR_Circuit} to appear within a model of an electromagnetic structure whose overall dimensions are electrically small. However, it occurs in this case because the \edit{electrically small} unit cell \edit{\sout{is able to fit a multi-wavelength} contains an electrically long} waveguide\edit{\sout{inside itself when the waveguide has been wrapped around itself multiple times}}. \edit{\sout{It is only because the unit cell is electrically small while the waveguide is electrically long that it makes sense for every small section of the waveguide to be mutually coupled to the same inductor(s).}}

Finally, the effective permeability for the Swiss roll unit cell can be written as: 
\begin{equation}
\boxed{
\begin{aligned}
    \label{eqn:LongSR_muZZ}
    \mu_{zz}& = 1-\frac{L_o-L_i}{L_{c}}+\frac{2Z_{0}}{\omega L_c}\tan\left( \frac{\gamma}{2} \right) \\
        & -\frac{1}{j\omega L_{c}}\frac{\left( j\omega L_{i}+jZ_{0}\tan\left( \frac{\gamma}{2} \right) \right)^{2}}{j\omega L_{i}+R_{i}-jZ_{0}\cot(\gamma)}.
\end{aligned}
}
\end{equation}

\subsection{Comparing the Model with Full-Wave Simulation}
\label{subsec:Full_Wave_Comparison}
To verify the model, we can compare the analytic dispersion diagram of the Swiss roll metamaterial with full-wave simulations. Since the Swiss roll metamaterial is approximately uniaxial (isotropic in the $xy$ plane), the dispersion diagram for propagation in the transverse plane ($k_z=0$) is simply
\begin{equation}
    \label{eqn:RES_DispersionEqn}
    k_t= \frac{\omega}{c} \sqrt{\mu_{zz}\epsilon_t},
\end{equation}
where $\epsilon_t$ is the transverse relative permittivity. 

In order to estimate the transverse permittivity, consider that the dominant effect of the local electric field within the unit cell is to polarize the scatterer by forcing charges to opposite locations on the outside of the scatterer. A good estimate of the transverse permittivity can therefore be found by treating the Swiss roll as a solid PEC cylinder. It is straightforward to find the scattered field from this cylinder. Averaging the total electric field over the unit cell results in the transverse relative permittivity: 
\begin{equation}
    \label{eqn:RES_epsTransverse}
    \epsilon_t = \frac{1}{1-\frac{\pi a_o^2}{w_c^2}},
\end{equation}
where $a_o$ is the outer radius of the Swiss roll. This tells us that the permittivity is approximately frequency independent and depends only on the fill fraction of the unit cell. 

Combining the transverse permittivity with the permeability (\ref{eqn:LongSR_muZZ}) then gives the transverse wavenumber as a function of frequency according to (\ref{eqn:RES_DispersionEqn}). Fig. \ref{fig:Results}a compares the analytic dispersion diagram with the dispersion computed by using the eigenmode solver in Ansys HFSS. For these results, the unit cell has side length $w_c=\frac{\lambda}{10}$ at 300 MHz, while the diameter of the Swiss roll is $2a_o=\frac{\lambda}{30}$ and the spiral waveguide is a full wavelength long. 

There is excellent agreement between the analytic model and the HFSS results. The dispersion diagram closely follows the light line, except in the vicinity of resonances within the Swiss roll that cause strong flat bands to appear. Narrow stop bands in between these resonances are frequencies over which the effective permeability of the unit cell is negative. 

To visualize these resonances, the axial magnetic field intensity at each resonance is plotted in Fig. \ref{fig:Results}b. At the lowest order resonance $(m=0)$, the waveguide is electrically short. This resonance arises when the inductance of the interior of the Swiss roll resonates with the capacitance of the short waveguide. The higher order resonances are characterized by standing wave patterns within the waveguide and are indexed according to the number of field maxima that occur within the guide. Note that \edit{\sout{in all cases,}} the scattered magnetic field is largely confined to the interior of the Swiss roll structure\edit{\sout{, justifying our neglect of nearest neighbor coupling in modeling the unit cell}}. 

\edit{\sout{The largest visible discrepancy between the full-wave simulations and the analytic model is that the latter closely follows the light line away from resonance, while the simulations show a slightly slower phase velocity than free space. This effect is due to coupling between neighboring unit cells that is not accounted for in our model.}}

\section{Discussion and Summary}
We developed an intuitive model for the permeability of the Swiss roll metamaterial which is valid even for lossy, finite thickness conductors and for any number of turns around the axis of the spiral. This model was derived by mapping the problem of finding the homogenized permeability to the equivalent problem of finding an effective inductance for the unit cell. Recognizing that independent electric currents can flow on either side of the conductor, the Swiss roll can be seen as a parallel plate waveguide wrapped around itself. Current entering or exiting the ends of this spiral waveguide flows around the interior or exterior perimeters of the Swiss roll. By viewing these currents as if flowing in free space, it is straightforward to intuit the associated electric and magnetic field distributions. Satisfying the boundary conditions imposed by the conductor simply requires applying the integral form of Faraday's law to appropriately chosen paths. \edit{\sout{Magnetic flux flowing through such a closed loop due to currents other than those bounding the loop results in mutual inductance between portions of the structure.}}

The resulting system of equations maps directly to the mesh equations of the equivalent circuit model of the Swiss roll (Fig. \ref{fig:LongSR_Circuit}). Currents in the circuit model correspond to the currents flowing on the Swiss roll. Mutual inductance plays an important roll in the behavior of the circuit model. Indeed, the effective permeability of the Swiss roll (and the other geometries considered) gets its form from impedance inversion through the mutual inductance between the scatterer and rest of the unit cell. 

Importantly, there is a distributed mutual inductance between the spiral waveguide and currents flowing \edit{\sout{exterior to the Swiss roll}outside the waveguide}. This creates what we have called a flux-coupled waveguide. Such a guide is somewhat unusual, but it is more than a scientific curiosity. For example, nonlinear flux-coupled transmission lines based on Josephson junctions have been proposed as broadband isolators for cryogenic RF systems (e.g., quantum computers) \cite{2025caloz_Isolator}. 

It is also worth noting that the resonant flat bands characterizing the Swiss roll metamaterial are desirable as properties of a photonic time crystal. As noted in \cite{1999pendry_SwissRoll}, the strong resonances cause significant field enhancement inside the Swiss roll. This can be used to activate nonlinearities, driving time-varying material properties. Flat bands from the sharp resonances (Fig. \ref{fig:Results}a) have recently been shown to allow for broadband momentum bandgaps, even with small modulation depths \cite{2025garg_MomBandGap}.

Ten years after the original publication of the Swiss roll negative permeability metamaterial, a follow-on numerical study \cite{2009demetriadou_SwissRollSims} noted both the presence of higher order resonances in the Swiss roll medium and the difficulty of modeling the permeability of the Swiss roll. Here, at last, we have provided an analytic model of the Swiss roll that predicts the effective permeability of the medium -- even the higher order resonances -- while clarifying the equivalent circuit perspective on the Swiss roll medium.

\appendices
\section{Geometric Parameters for the Swiss Roll Unit Cell}
\label{Appendix_Geometry}
The Swiss roll is treated as generated by an Archimedean spiral, parameterized in cylindrical coordinates by: 
\begin{equation}
    \label{eqn:appA_archSpiral}
    \rho(\phi)=a_o-T \frac{\phi}{2\pi},
\end{equation}
where $a_o$ is the outer radius of the Swiss roll and $T$ is the thickness of each turn. The azithumal angle varies over the interval $\phi\in[0,2\pi \mathcal{N}]$, so that $\mathcal{N}-1$ is the number of times the spiral waveguide wraps fully around its axis. Given this parameterization, the length of the outer and inner perimeters of the Swiss roll can be approximated by: 
\begin{equation}
    \ell_{o,i} \approx 2\pi \left(a_{o,i}-\frac{T}{2}\right),
\end{equation}
which holds if the pitch of the turns is very small, i.e., $\frac{T}{a_o}\ll 1$. The parameter $t_c=T-g$ is the thickness of the conductor and $a_i$ is the inner radius of the Swiss roll, given by:
\begin{equation}
    a_i \approx a_o-\mathcal{N}T-t_c.
\end{equation} 
The length of the spiral waveguide is approximately given as:
\begin{equation}
    d \approx 2\pi (\mathcal{N}-1)\left( a_o - \frac{\mathcal{N}T+t_c}{2} \right).
\end{equation}

The area enclosed by the outer and inner perimeters of the Swiss roll are approximately: 
\begin{equation}
  A_{o,i}  \approx \pi a_{o,i}^2\left( 1-\frac{T}{a_{o,i}} \right).
\end{equation}

\section{Diagonalizing a Symmetric, Tridiagonal Toeplitz Matrix}
\label{Appendix_Toeplitz}
Tridiagonal Toeplitz matrices are special, as they are one of the few patterned matrices that can be analytically diagonalized regardless of the size of the matrix. Any eigenvector $\mathbf{\hat{I}}$ of (\ref{eqn:Toeplitz_Matrix}) satisfies: 
\begin{equation}
    \label{eqn:EigProblem_Matrix_Form}
    z\mathbf{\hat{I}}=\mathbf{Z}_{22}\mathbf{\hat{I}}.
\end{equation}
If we consider row $k$ of the eigenproblem, we obtain: 
\begin{equation}
    \label{eqn:DiffEqn_Prototype}
    z I_{k}=aI_{k+1}+bI_{k}+aI_{k-1}, \quad k\in \{ 2,\dots,M-2 \}
\end{equation}
which does not hold for the first and final rows of the eigenproblem, since these rows of the matrix only have two nonzero entries. However, if we treat $\{ I_{k} \}$ as a sequence, then it is possible to rewrite the eigenproblem as the linear, homogeneous, second-order difference equation with constant coefficients: 
\begin{equation}
    \label{eqn:Difference_Eqn}
    0=aI_{k+1}+(b-z)I_{k}+aI_{k-1}, \quad  k\in \{ 1,\dots,M-1 \}
\end{equation}
with the boundary conditions $I_{0}=I_{M}=0$. To solve this difference equation, we assume a solution of the form $I_{k}=x^{k}$. Substituting this into the difference equation yields a quadratic equation for $x$,
\begin{equation}
    \label{eqn:Quadratic_for_DiffEqn}
    ax^{2}+(b-z)x+a=0.
\end{equation}
The roots of this equation are simply,
\begin{equation}
    \label{eqn:Roots_Quadratic_Formula}
    x_{\pm}=\frac{z-b}{2}\pm \frac{1}{2}\sqrt{ (z-b)^{2}- 4a^{2}}.
\end{equation}
Solving the difference equation requires a superposition of two basic solutions $I_{k}=c_{1}x_{+}^{k}+c_{2}x_{-}^{k},$ where the constants $c_{1,2}$ must be chosen so that the boundary conditions are satisfied. According to the first boundary condition: 
\begin{equation}
    \label{eqn:DiffEqn_BC_0}
    I_{0}=c_{1}+c_{2}=0.
\end{equation}
Applying this to the second boundary condition yields:
\begin{equation}
\label{eqn:DiffEqn_BC_M}
    I_{M}=c_{1}(x_{+}^{M}-x_{-}^{M})=0,
\end{equation}
which implies that the ratio of the roots can be any of the $M$ roots of unity: 
\begin{equation}
    \label{eqn:DiffEqn_Ratio_of_Roots}
    \frac{x_{+}}{x_{-}}=e^{j \frac{2\pi m}{M}}.
\end{equation}
It is simple to verify that the product of the roots is equal to unity. In fact, this must be the case, since the quadratic equation (\ref{eqn:Quadratic_for_DiffEqn}) is invariant under the substitution $x \rightarrow \frac{1}{x}$, so that if $x_{+}\neq0$ is a root, then $x_{-}=\frac{1}{x_{+}}$ must also be a root. Consequently, the eigenvectors from the solutions to the difference equation are: 
\begin{equation}
    \label{eqn:Eigenvectors}
    [\mathbf{\hat{I}}_{m}]_{k}=I_{k,m}=\sqrt{ \frac{2}{M} }\sin\left( \frac{\pi km}{M} \right).
\end{equation}
The eigenvalue associated with the $m^{th}$ eigenvector is given by: 
\begin{equation}
    \label{eqn:Eivenvalues}
    z_{m} =a+2b\cos\left( \frac{m\pi}{M} \right)
\end{equation}
It is straightforward to show that the eigenvectors $\{ \mathbf{\hat{I}}_{m} \}$ form an orthonormal set. Therefore, the matrix $\mathbf{Z}_{22}$ is unitarily diagonalized as: 
\begin{equation}
    \mathbf{Z}_{22}=\mathbf{U}\mathbf{D}\mathbf{U},
\end{equation}
where $\mathbf{U}=\begin{bmatrix} \hat{\mathbf{I}}_{1} & \hat{\mathbf{I}}_{2} & \cdots & \hat{\mathbf{I}}_{N-2}\end{bmatrix}$ is the matrix of eigenvectors, which is unitary and transpose symmetric, and $\mathbf{D}$ is the diagonal matrix of eigenvalues: 
\begin{equation}
    \mathbf{D}_{\ell,m}=\
        \begin{cases}
        z_{m}  & \ell=m \\
        0  & \ell\neq m
        \end{cases}.
\end{equation}

\ifCLASSOPTIONcaptionsoff
  \newpage
\fi

\bibliographystyle{IEEEtran}
\bibliography{ref}

@article{1999pendry_SwissRoll,
  title = {{M}agnetism from Conductors and Enhanced Nonlinear Phenomena},
  author = {Pendry, J.B. and Holden, A.J. and Robbins, D.J. and Stewart, W.J.},
  year = 1999,
  month = nov,
  journal = {IEEE Trans. on Microw. Theory and Tech.},
  volume = {47},
  number = {11},
  pages = {2075--2084},
  issn = {1557-9670},
  doi = {10.1109/22.798002}
  }

@article{2001wiltshire_SwissRollMRI,
  title = {Microstructured Magnetic Materials for {RF} Flux Guides in Magnetic Resonance Imaging},
  author = {Wiltshire, M. C. K. and Pendry, J. B. and Young, I. R. and Larkman, D. J. and Gilderdale, D. J. and Hajnal, J. V.},
  year = 2001,
  month = feb,
  journal = {Science},
  volume = {291},
  number = {5505},
  pages = {849--851},
  publisher = {American Association for the Advancement of Science}
 }

@inproceedings{2003wiltshire_SwissRollSeminar,
  title = {``{{Swiss}} Roll" Metamaterials - an Effective Medium with Strongly Negative Permeability},
  booktitle = {{{IEE Seminar}} on {{Metamaterials}} for {{Microwave}} and ({{Sub}}) {{Millimetre Wave Applications}}: {{Photonic Bandgap}} and {{Double Negative Designs}}, {{Components}} and {{Experiments}} 2003},
  author = {Wiltshire, M.C.K. and Pendry, J.B. and Hajnal, J.V. and Edwards, D.J.},
  year = 2003,
  month = nov,
  pages = {13/1-13/10}
  }

@article{2004wiltshire_1DArraySwissRolls,
  title = {Experimental and Theoretical Study of Magneto-Inductive Waves Supported by One-Dimensional Arrays of ``{S}wiss Rolls''},
  author = {Wiltshire, M. C. K. and Shamonina, E. and Young, I. R. and Solymar, L.},
  year = 2004,
  month = apr,
  journal = {J. of Appl. Phys.},
  volume = {95},
  number = {8},
  pages = {4488--4493},
  issn = {0021-8979},
  doi = {10.1063/1.1687036}
 }

@article{2007zolla_SwissRollSims,
  title = {Swiss Roll Lattices: {N}umerical and Asymptotic Modeling},
  author = {Zolla, F. and Nicolet, A. and Guenneau, S.},
  year = 2007,
  month = oct,
  journal = {Waves Random Complex Media},
  volume = {17},
  number = {4},
  pages = {571--579},
  publisher = {Taylor \& Francis},
  issn = {1745-5030},
  doi = {10.1080/17455030701504350}
  }

@article{2009demetriadou_ChiralSwissRoll,
  title = {Extreme Chirality in {{Swiss}} Roll Metamaterials},
  author = {Demetriadou, A and Pendry, J B},
  year = 2009,
  month = aug,
  journal = {J. Phys. Condens. Matter},
  volume = {21},
  number = {37},
  pages = {376003},
  issn = {0953-8984},
  doi = {10.1088/0953-8984/21/37/376003}
 }

@article{2009demetriadou_SwissRollSims,
  title = {Numerical Analysis of {{Swiss}} Roll Metamaterials},
  author = {Demetriadou, A and Pendry, J B},
  year = 2009,
  month = jul,
  journal = {J. Phys. Condens. Matter},
  volume = {21},
  number = {32},
  pages = {326006},
  issn = {0953-8984},
  doi = {10.1088/0953-8984/21/32/326006},
  urldate = {2025-09-20}
  }

@article{2015chen_SwissRollAbsorber,
  title = {Microwave Absorber Based on Permeability-near-Zero Metamaterial Made of {{Swiss}} Roll Structures},
  author = {Chen, Ke and Jia, Nan and Sima, Boyu and Zhu, Bo and Zhao, Junming and Feng, Yijun and Jiang, Tian},
  year = 2015,
  month = oct,
  journal = {J. Phys. D: Appl. Phys.},
  volume = {48},
  number = {45},
  pages = {455304},
  publisher = {IOP Publishing},
  issn = {0022-3727},
  doi = {10.1088/0022-3727/48/45/455304}
 }

@article{1996pendry_NegEpsWireMedium,
  title = {Extremely Low Frequency Plasmons in Metallic Mesostructures},
  author = {Pendry, J. B.},
  year = 1996,
  journal = {Phys. Rev. Lett.},
  volume = {76},
  number = {25},
  pages = {4773--4776}
}

@article{2000pendry_NegIndPerfectLens,
  title = {Negative Refraction Makes a Perfect Lens},
  author = {Pendry, J. B.},
  year = 2000,
  journal = {Phys. Rev. Lett.},
  volume = {85},
  number = {18},
  pages = {3966--3969}
}

@article{2000smith_DoubleNegative,
  title = {Composite Medium with Simultaneously Negative Permeability and Permittivity},
  author = {Smith, D. R.},
  year = 2000,
  journal = {Phys. Rev. Lett.},
  volume = {84},
  number = {18},
  pages = {4184--4187}
}

@article{1968veselago_DoubleNegative,
  title = {The Electrodynamics of Substances with Simultaneously Negative Values of {$\epsilon$} and {$\mu$}},
  author = {Veselago, Viktor G.},
  year = 1968,
  month = apr,
  journal = {Sov. Phys. Usp.},
  volume = {10},
  number = {4},
  pages = {509},
  publisher = {IOP Publishing},
  issn = {0038-5670},
  doi = {10.1070/PU1968v010n04ABEH003699}
}

@article{2004grbic_DiffractionLimit,
  title = {Overcoming the Diffraction Limit with a Planar Left-Handed Transmission-Line Lens},
  author = {Grbic, Anthony},
  year = 2004,
  journal = {Phys. Rev. Lett.},
  volume = {92},
  number = {11},
  doi = {10.1103/PhysRevLett.92.117403}
}

@article{1998pendry_WirePlasmaExpt,
  title = {Low Frequency Plasmons in Thin-Wire Structures},
  author = {Pendry, J. B. and Holden, A. J. and Robbins, D. J. and Stewart, W. J.},
  year = 1998,
  month = jun,
  journal = {J. Phys. Condens. Matter},
  volume = {10},
  number = {22},
  pages = {4785},
  issn = {0953-8984},
  doi = {10.1088/0953-8984/10/22/007}
 }

@incollection{2001harrington_EquivalencePrinciple,
  title = {Some Theorems and Concepts},
  booktitle = {Time-Harmonic Electromagnetic Fields},
  author = {Harrington, Roger F.},
  year = 2001,
  pages = {95--142},
  publisher = {IEEE},
  doi = {10.1109/9780470546710.ch3}
 }

@article{2006smith_FieldAveraging,
  title = {Homogenization of Metamaterials by Field Averaging (Invited Paper)},
  author = {Smith, David R. and Pendry, John B.},
  year = 2006,
  month = mar,
  journal = {JOSA B},
  volume = {23},
  number = {3},
  pages = {391--403},
  publisher = {Optica Publishing Group},
  issn = {1520-8540},
  doi = {10.1364/JOSAB.23.000391}
 }

@article{2014grbic_ShieldedLoopRes,
  title = {Planar Shielded-Loop Resonators},
  author = {Tierney, Brian B. and Grbic, Anthony},
  year = 2014,
  month = jun,
  journal = {IEEE Trans. Antennas Propag.},
  volume = {62},
  number = {6},
  pages = {3310--3320},
  issn = {1558-2221},
  doi = {10.1109/TAP.2014.2314305}
}

@article{1981hardy_SRRs,
  title = {Split-ring Resonator for Use in Magnetic Resonance from 200--2000 {{MHz}}},
  author = {Hardy, W. N. and Whitehead, L. A.},
  year = 1981,
  month = feb,
  journal = {Review of Scientific Instruments},
  volume = {52},
  number = {2},
  pages = {213--216},
  issn = {0034-6748},
  doi = {10.1063/1.1136574}
}

@book{1952schelkunoff_AntennasTheoryPractice,
  title = {Antennas: {{Theory}} and Practice},
  author = {Schelkunoff, Sergei A. and Friis, Harald T.},
  year = 1952,
  publisher = {Wiley},
  address = {New York}
}

@article{2025garg_MomBandGap,
  title = {Expanding Momentum Bandgaps in Photonic Time Crystals through Resonances},
  author = {Wang, X. and Garg, P. and Mirmoosa, M. S. and Lamprianidis, A. G. and Rockstuhl, C. and Asadchy, V. S.},
  year = 2025,
  month = feb,
  journal = {Nat. Photon.},
  volume = {19},
  number = {2},
  pages = {149--155},
  publisher = {Nature Publishing Group},
  issn = {1749-4893},
  doi = {10.1038/s41566-024-01563-3}
}

@article{1966_yeeGrid,
  title = {Numerical Solution of Initial Boundary Value Problems Involving {M}axwell's Equations in Isotropic Media},
  author = {Yee, Kane},
  year = 1966,
  month = may,
  journal = {IEEE Trans. Antennas Propag.},
  volume = {14},
  number = {3},
  pages = {302--307},
  issn = {1558-2221},
  doi = {10.1109/TAP.1966.1138693}
 }

@article{2019nakata_SelfDuality,
  title = {Geometric Structure behind Duality and Manifestation of Self-Duality from Electrical Circuits to Metamaterials},
  author = {Nakata, Yosuke and Urade, Yoshiro and Nakanishi, Toshihiro},
  year = 2019,
  month = nov,
  journal = {Symmetry},
  volume = {11},
  number = {11},
  pages = {1336},
  publisher = {Multidisciplinary Digital Publishing Institute},
  issn = {2073-8994},
  doi = {10.3390/sym11111336}
 }

@misc{2025caloz_Isolator,
  title = {Broadband Magnetless Isolator Using Adiabatic Flux Modulation},
  author = {Demarets, M. and Vadiraj, A. M. and Caloz, C. and Greve, K. De},
  year = 2025,
  month = sep,
  number = {arXiv:2509.24551},
  eprint = {2509.24551},
  primaryclass = {quant-ph},
  publisher = {arXiv},
  doi = {10.48550/arXiv.2509.24551},
  urldate = {2025-11-26}
  }

@article{1953brown_IndexLessThan1,
  title={Artificial dielectrics having refractive indices less than unity},
  author={Brown, John},
  journal={Proc. IEE - Part IV},
  volume={100},
  number={5},
  pages={51--62},
  year={1953},
  publisher={IET}
}

@article{1955brown_ArtificialDielectricsCM,
  title = {The Properties of Artificial Dielectrics at Centimetre Wavelengths},
  author = {Brown, John and Jackson, Willis},
  year = 1955,
  month = jan,
  journal = {Proc. IEE - Part B},
  volume = {102},
  number = {1},
  pages = {11--16}
}

@article{1962rotman_ArtificialPlasma,
  title = {Plasma Simulation by Artificial Dielectrics and Parallel-Plate Media},
  author = {Rotman, W.},
  year = 1962,
  month = jan,
  journal = {IRE Trans. Antennas Propag.},
  volume = {10},
  number = {1},
  pages = {82--95}
}

@article{model1955propagation,
  title={Propagation of plane electromagnetic waves in a space which is filled with plane parallel grids},
  author={Model, A M},
  journal={Radiotekhnika},
  volume={10},
  pages={52--57},
  year={1955}
}

\end{document}